\documentclass[letter]{aa}  

\usepackage{graphicx}
\usepackage{fixltx2e}
\usepackage[utf8]{inputenc}
\everymath={\rm}
\usepackage{txfonts}
\usepackage{natbib}
\usepackage{dashbox,framed,color,ocg-p}
\usepackage[x11names]{xcolor}
\newcommand{\ToggleLayer}[2]{%
  \leavevmode
  \pdfstartlink user {
    /Subtype /Link
    /Border [0 0 0]%
    /A <<
      /S/JavaScript
      /JS (
         var aOCGs = this.getOCGs(), Layer;
         var Layers = "#1".split(","), Active = -1, i, l;
         for (l=0; l<Layers.length; l++) {
           Layer = Layers[l];
           for (i=0; aOCGs && i<aOCGs.length; i++) {
             if (aOCGs[i].state && aOCGs[i].name == Layer) {
               Active = l;
               aOCGs[i].state = false;
             }
           }
           if (Active >= 0) break;
         }
         if (Active == -1) {
           for (l=0; l<Layers.length; l++) {
             if (Layers[l] == "") Active = l;
           }
         }
         Active = Active + 1;
         if (Active == Layers.length) Active = 0;
         Layer = Layers[Active];
         for (i=0; aOCGs && i<aOCGs.length; i++) {
           if (aOCGs[i].name == Layer) aOCGs[i].state = true;
         }
      )
    >>
  }#2%
  \pdfendlink
}
\newcommand{\cdbox}[1]{%
  \colorlet{currentcolor}{.}%
  {\color{DodgerBlue2}%
    \dbox{\color{currentcolor}#1}}%
}

\newcommand{\cdhob}{C$^{18}$O }

\newcommand{\ndhp}{N$_2$H$^+$}
\newcommand{\ndhpb}{N$_2$H$^+$ }

\newcommand{\cc}{cm$^{-3}$}
\newcommand{\ccb}{cm$^{-3}$ }
\newcommand{\sqc}{cm$^{-2}$}
\newcommand{\sqcb}{cm$^{-2}$ }

\newcommand{\mjy}{MJy\,sr$^{-1}$}
\newcommand{\Av}{A$_{\mathrm V}$}
\newcommand{\Avb}{A$_{\mathrm V}$ }

\newcommand{\pdix}[1]{$\times$\,10$^{#1}$}
\newcommand{\pdixb}[1]{$\times$\,10$^{#1}$ }

\newcommand{\mic}{$\mu$m}

\begin{document}

   \title{Can we trace very cold dust from its emission alone ?
\thanks{see Appendix \ref{AppB} for institutional acknowledgements}%
}
   \author{L. Pagani           \inst{1,2}
           \and
           C. Lef\`{e}vre\inst{1,2}
          \and M. Juvela \inst{3}
          \and
V.-M. Pelkonen \inst{3,4}
\and
	F. Schuller \inst{5}
}
   \offprints{L.Pagani}

 \institute{ LERMA, Observatoire de
  Paris, PSL Research University, CNRS, UMR 8112, F-75014 Paris, France\\
\email{laurent.pagani@obspm.fr}
\and
Sorbonne Universit\'es, UPMC Univ. Paris 6, UMR 8112, LERMA, F-75005, Paris, France 
\and
Department of Physics, P.O.Box 64, FI-00014, University of Helsinki,
Finland, {\em mika.juvela@helsinki.fi}                                
\and
Finnish Centre for Astronomy with ESO (FINCA), University of Turku,
V\"ais\"al\"antie 20, FI-21500 Piikki\"o, Finland
\and
European Southern Observatory, Alonso de Córdova 3107, Vitacura, Casilla 19001, Santiago de Chile, Chile
           }

   \date{Received 02/10/2014; accepted 23/12/2014}

 
  \abstract
   {Dust is a good tracer of cold dark clouds but its column density is difficult to quantify.}
   {We want to check whether the far-infrared and submillimeter high--resolution data from \textit{Herschel} SPIRE and PACS cameras combined with ground-based telescope bolometers allow us to retrieve the whole dust content of cold dark clouds.}
   {We compare far-infrared and submillimeter emission across L183 to the 8 \mic\ absorption map from Spitzer data and fit modified blackbody functions towards three different positions.}
   {We find that none of the \textit{Herschel} SPIRE channels follow the cold dust profile seen in absorption. Even the ground-based submillimeter telescope observations, although more closely  following the absorption profile, cannot help to characterize the cold dust without external information such as the dust column density itself. The difference in dust opacity can reach up to a factor of ~3  in prestellar cores of high extinction.}
   {In dark clouds, the amount of very cold dust cannot be measured from its emission alone. In particular, studies of dark clouds based  only on \textit{Herschel} data can miss a large fraction of the dust content. This has an impact on core and filament density profiles, masse and stability estimates.}



   \maketitle
%

\section{Introduction}
Dark clouds are the places where stars form. We try to follow the different steps that lead from a low--density cloud to  
main--sequence star. Some steps have been clearly identified like the Class 0 to Class III protostar evolution \citep{Lada1987,Andre1993} but,  
the early phases, such as the (formation of a prestellar core (PSC), the collapse of this core, are still not well understood. Studies of clouds and 
of PSCs  first attempt to determine their mass. This is not a simple task. The first difficulty is to know the distance of the object, and the 
second is to trace the column density of the material itself. The main components, H$_2$ and He are not directly visible, with the  exception of a little H$_2$ seen in 
absorption in the UV at the cloud edges, and surrogate tracers are needed. Molecules are not good tracers in general because 
CO, which is the standard tracer, is depleted in the PSCs \citep{Lemme1995,Willacy1998,Tafalla2002,Pagani2005,BradyFord:2011ke}.  
Tracers of the PSCs like NH$_3$ or \ndhpb do not extend beyond the PSCs themselves, and contrary to CO, their peak
abundance is variable. Detailed radiative transfer models are therefore needed to retrieve the H$_2$ + He densities via 
the modelling of their collisions with the tracers (followed by integration along the line of sight to obtain the total column density). On the other hand, dust is traceable from the edge of the clouds to the 
centre. Its relative abundance to H$_2$ is not accurately known but is thought to be in the range $\sim$1/130--1/100
in the Milky Way \citep{Flower2005,Compiegne:2011jf}. It can 
be traced either by its extinction of background stars \citep{Wolf1923,Bok1956,Bok1973} with a higher efficiency in the near-infrared 
\citep[NIR,][]{Lada1994,Lombardi:2001bka,Lombardi:2009hd} or of background diffuse light in the mid-infrared 
\citep[MIR,][]{Bacmann2000}, by its scattering of the interstellar radiation field (ISRF)  in the NIR and MIR 
\citep{Lehtinen:1996ti,Juvela2006,Lef`evre2014}, or by its emission in the far-infrared (FIR) \citep{Ward-Thompson1994}.
However, all these methods have difficulties. The use of background stars do not allow to reach a high spatial resolution outside the 
galactic plane and become absent typically when \Avb $\geq$ 40 mag. Scattering is a promising but difficult method, highly 
dependent on the type of grains (size distribution, extinction, albedo, phase function), on the background intensity, and on the 
ISRF anisotropy and strength (see \citealt{Lef`evre2014} for an exhaustive study).  Emission depends 
on the knowledge of the grain properties, spectral index and temperature (e.g. \citealt{Juvela:2012ex} for a discussion of the 
degeneracy of the problem). The advantage of dust emission measurements in the FIR and submillimetre domains is that it 
remains optically thin up to N(H$_2$)  $\sim$2\,\pdix{24}\,\sqc\ at 300 \mic\ and even higher at longer wavelengths. Dust 
observed in emission is therefore able to trace the whole cloud content from the edge to the centre. However, it remains difficult to convert this emission into the actual column density of dust.

Before the launch of \textit{Herschel} \citep{Pilbratt2010}, FIR measurements of dust emission in the range 
200--500 \mic\ were scarce \citep[e.g.][]{Stepnik2003} and many studies limited themselves to fit the 
dust in emission by building single  spectral energy distribution (SED) fits based on a mix of 
measurements shorter than 200 \mic\ and longer than 850 \mic. However, \citet{Pagani2004} 
showed  that in a dark cloud without embedded sources, the 200 \mic\ emission traces only the 
envelope of the cloud (\Avb $\leq$ 7.5 mag 
from the surface) and totally misses the bulk of the dust too cold to emit significantly at that 
wavelength. Hence, it could be deduced from that result that single SED fits  were not realistic because 
short wavelengths are mostly tracing emission from the envelope and long wavelengths emission from the core. There was no 
connection between the two sides of the (modified) blackbody curve and therefore no physical 
meaning to that fit.  The question arises again today now that  all wavelengths 
between 100 and 1200 \mic\ have been sampled in a number of clouds with several telescopes, 
including Spitzer \citep{Werner2004}, AKARI \citep{Murakami2007}, ground-based bolometers, and 
above all, \textit{Herschel} \citep{Pilbratt2010} and \textit{Planck} \citep{Tauber2010}, to find out if we can accurately 
retrieve the 
dust content  from fitting SEDs alone. Indeed, dust at 10 K is brightest at 300 \mic, while dust as cold 
as 6 K is brightest at 500 \mic. Both wavelengths are inside the range observed by SPIRE, the FIR 
camera of \textit{Herschel} \citep{Griffin2010} and not far from the two highest frequency channels of 
\textit{Planck}/HFI \citep{Lamarre2010}. Using \textit{Herschel} data, \citet{Nielbock:2012ih} achieved 
a successful temperature analysis of \object{B68} but they combined data from NIR to submm. 
\citet{Roy:2014bv} attempted to fit two cores (B68 and \object{L1689B}) using \textit{Herschel} data 
alone. They both find temperatures lower than provided by single blackbody fittings of the SEDs.
Some aspects of dust emission fitting of mock clouds have also been numerically explored by \citet{Malinen:2011eu}.

\section{Observations}

\begin{figure}[t]
\centering
\includegraphics[width=0.9\hsize]{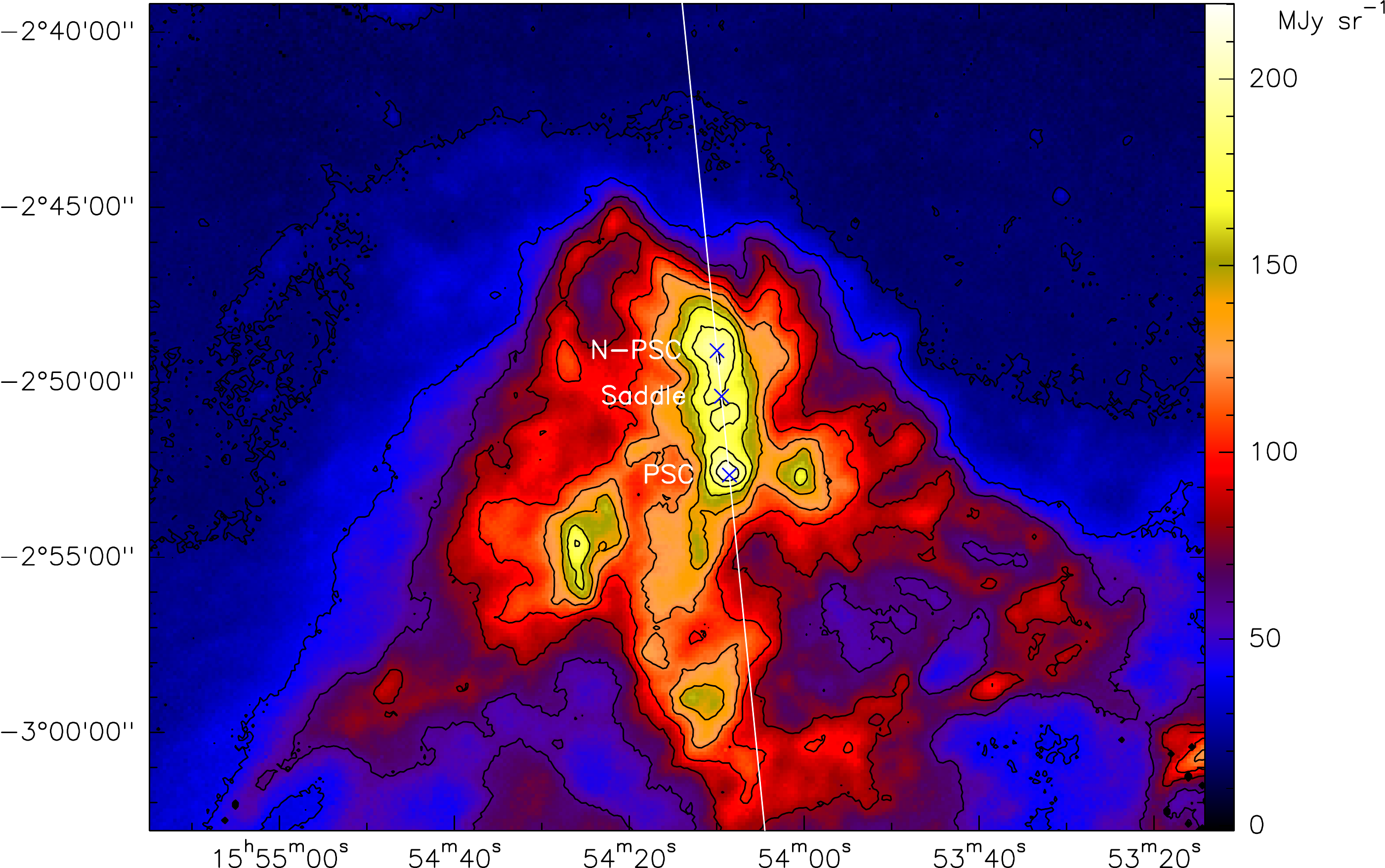}
\caption{L183 \textit{Herschel}/SPIRE map at 250 \mic. Contour levels every 20 \mjy. The cut through the region is traced as a $\sim$6\degr\ tilted line with respect to the vertical (equatorial North) direction.}
\label{fig:L183_CUT}
\end{figure}

This work is based on  maps of L183 taken at 8, 100, 160, 250, 350, 500, 870, and 1200 \mic. The 8 \mic\ data were taken  with Spitzer/IRAC, \citealt{Fazio:2004eb}, published in
\citealt{Steinacker2010}). \textit{Herschel}/PACS took the 100 and 160 \mic\ maps, and \textit{Herschel}/SPIRE the 250, 350, and 500 \mic\ maps
\citep{Pilbratt2010,Poglitsch2010,Griffin2010}. These maps are partly published in \citet[for the 250 \mic\
data]{2012A&A...541A..12J} and will be published in \citet{Montillaud-subm,Juvela-subm},  and Lef\`evre et al. (in
prep.). Lef\`evre et al.  will also publish two new APEX/LABOCA and IRAM-30m/MAMBO maps at 870 and 1200 \mic, respectively.
In this Letter, we only present a cut inclined eastward by $\sim$6\degr\ with respect to the vertical (Fig. 
\ref{fig:L183_CUT}). It is adjusted to go through the main prestellar core (PSC) and the northern prestellar core (N-PSC, Figs. \ref{fig:L183_CUT}\&\ref{fig:L183_8_250_870um_3y}). The minimum in extinction between the two cores is named the saddle. Details about data reduction will be exposed in Lef\`evre et al. (in prep.). Shortly, \textit{Herschel} data  
acquired as part of the {\em Galactic cold cores} key project \citep{Juvela2010} were re-reduced with the \textit{Herschel} interactive 
processing environment (HIPE) v.12.1, using the official pipeline \citep{Ott2010}, including Scanamorphos v.23 for PACS data
\citep{Roussel2013}. They were subsequently colour-corrected, zeroed, and  rescaled using IRIS 
\citep{MivilleDeschenes:2005fr} and \textit{Planck} data \citetext{Juvela-subm}. {The calibration error is a small fraction of the background subtraction uncertainty  which we estimate at 10\% for SPIRE and 15\% for PACS.} We reduced LABOCA data  using BoA\footnote{http://www.eso.org/sci/activities/apexsv/labocasv.html},  and MAMBO data using MOPSIC\footnote{http://www.iram.es/IRAMES/mainWiki/CookbookMopsic}. Both LABOCA 
and MAMBO data lack the large scale, low surface brightness part of the cloud because of the observing method of MAMBO (in-source OFF 
subtraction) and the data reduction method for LABOCA, which requires strong filtering to remove  sky fluctuations. 
To recover most of the lost signal, we 
combined these data with \textit{Planck} data that are colour-corrected and interpolated to match the LABOCA and MAMBO filter 
band-passes. For MAMBO, the correction is no more than 10\%, while for LABOCA the correction amounts to 30 \%. {The final uncertainty is estimated to be $\sim$20\% for both.}

\section{Analysis}
\begin{figure}[t]
\centering
\includegraphics[width=0.9\hsize]{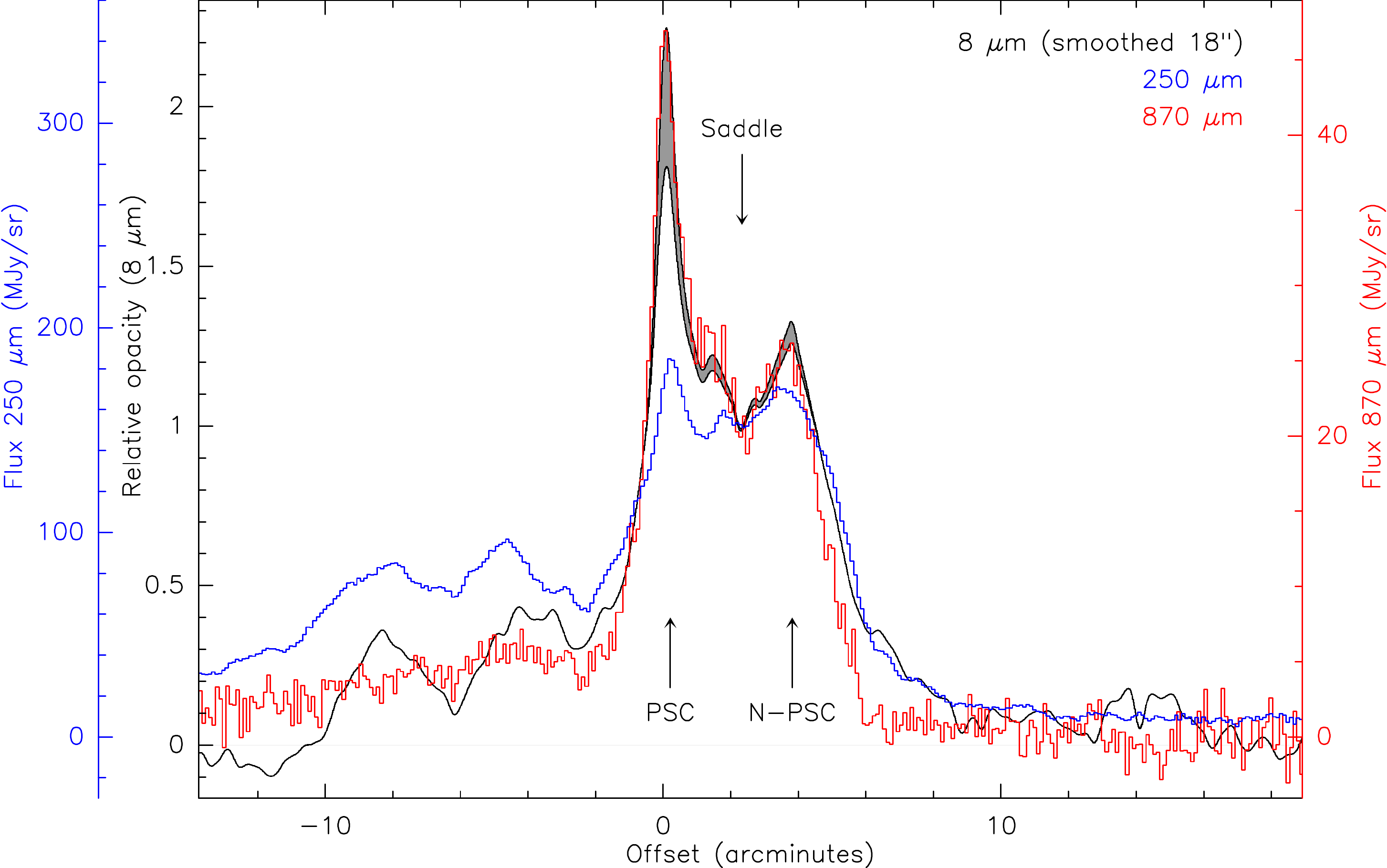}
\caption{L183 cut. The 8 \mic\ data have been smoothed to 18\arcsec\ resolution to match both the 
\textit{Herschel}/SPIRE 250 \mic\ and APEX/LABOCA 870 \mic\ common resolution. The 8 \mic\ 
absorption cut is converted to relative extinction (see text). The grey filling represents the possible 
range of 8 \mic\ ratio between main and northern peaks. All curves are aligned at the saddle point (at 
+2\arcmin33\arcsec offset).} 
\label{fig:L183_8_250_870um_3y}
\vspace*{-0.4cm}
\end{figure}

Figure \ref{fig:L183_8_250_870um_3y} shows the extinction profile along the cut (as defined in Fig. 
\ref{fig:L183_CUT}), derived from the 8 \mic\ data degraded to the resolution of, and compared with, 
the 250 \mic\ and the 870 \mic\ cuts. The 8 \mic\ opacity values 
are only a qualitative representation of the expected extinction. It is based on a previous estimate by \cite{Pagani2004}, on the necessity to 
reach zero extinction at the edges of the cloud, and on the estimated ratio in column density between the PSC and 
the N-PSC, which is 1.5--2 at the MAMBO 12\arcsec\ resolution (after smoothing 
to 18\arcsec, the  ratio is in the range 1.4--1.7). The extinction is represented as a range between these two 
ratio values. These values are only indicative since the 8 \mic\ absorption map  is in fact strongly 
contaminated by scattered light and cannot be safely converted to an opacity map without a 3D model 
taking the scattering due to micron-sized grains  into account \citep{Steinacker2010, 
Pagani2010a,Lef`evre2014}. A better 
estimate of the extinction based on NIR and MIR absorption and scattering, including the 8 \mic\  
map, 
will be presented in Lef\`evre et al. (in prep.). 

\begin{figure}[t!]
\centering
 \begin{ocg}{fig:PACSr}{fig:PACSr}{0}%
     \includegraphics[width=0.8\hsize]{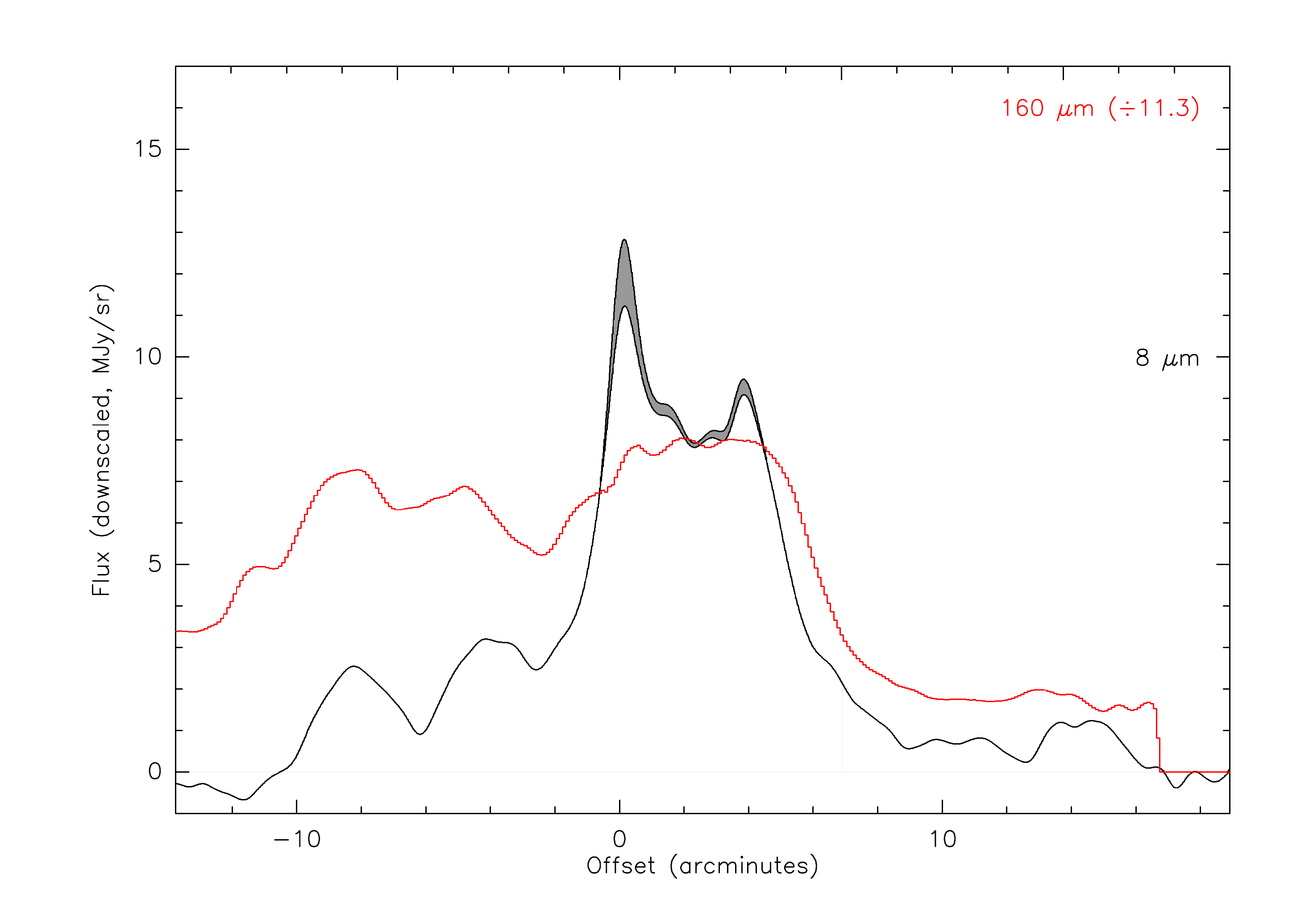}%
   \end{ocg}%
   \hspace{-0.8\hsize}%
 \begin{ocg}{fig:SPIRES}{fig:SPIRES}{0}%
     \includegraphics[width=0.8\hsize]{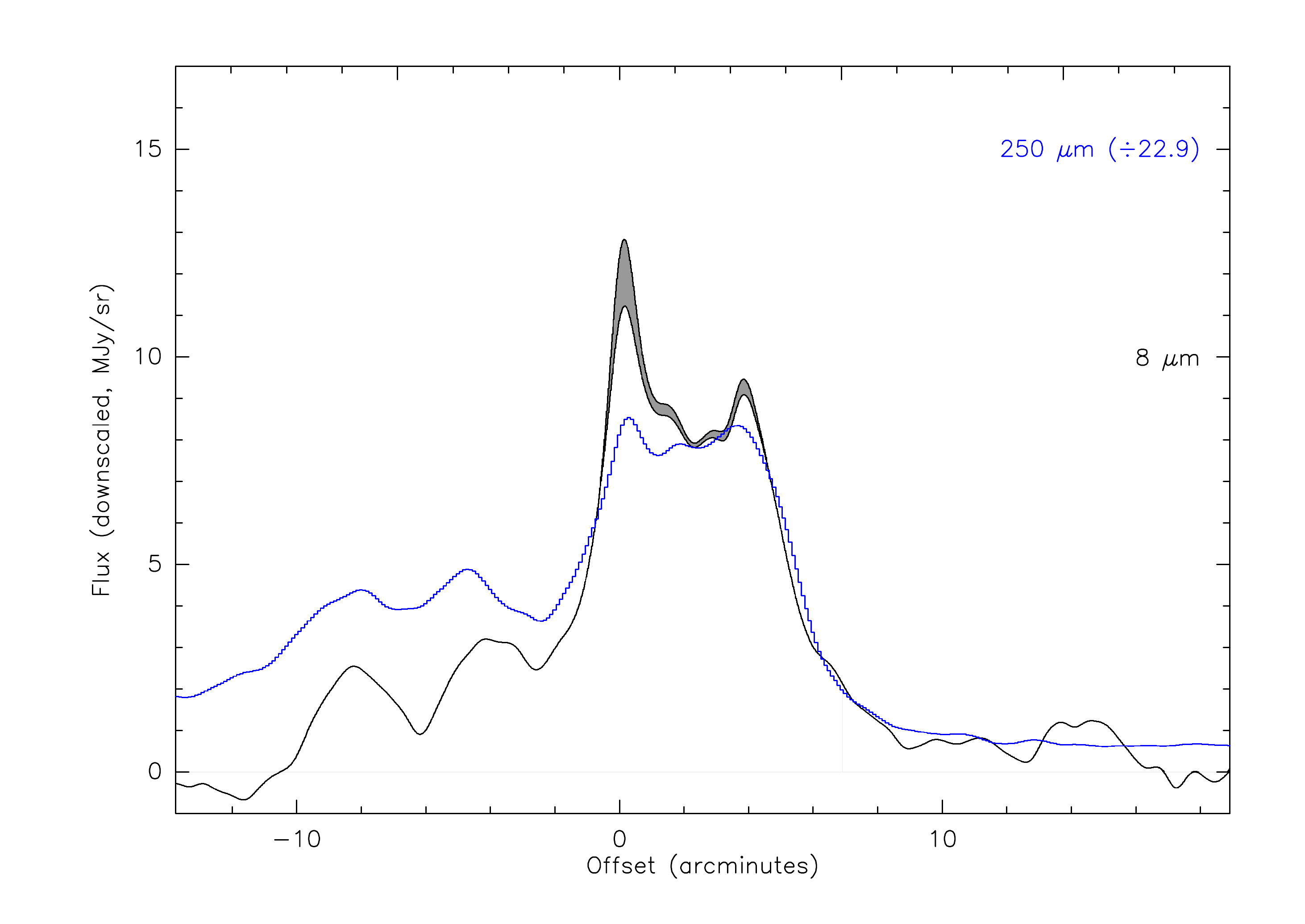}%
   \end{ocg}%
   \hspace{-0.8\hsize}%
 \begin{ocg}{fig:SPIREM}{fig:SPIREM}{0}%
     \includegraphics[width=0.8\hsize]{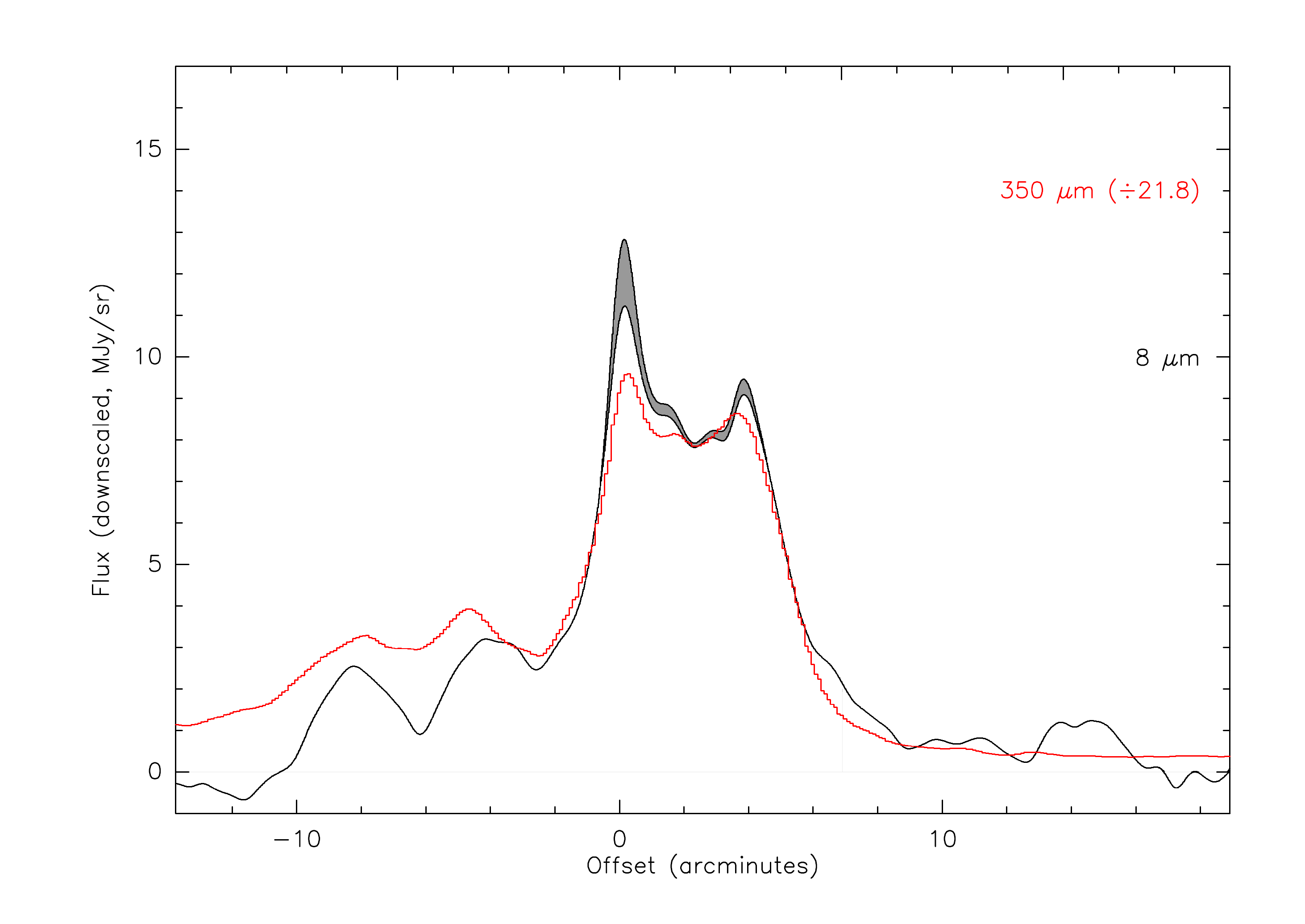}%
   \end{ocg}%
   \hspace{-0.8\hsize}%
 \begin{ocg}{fig:SPIREL}{fig:SPIREL}{0}%
     \includegraphics[width=0.8\hsize]{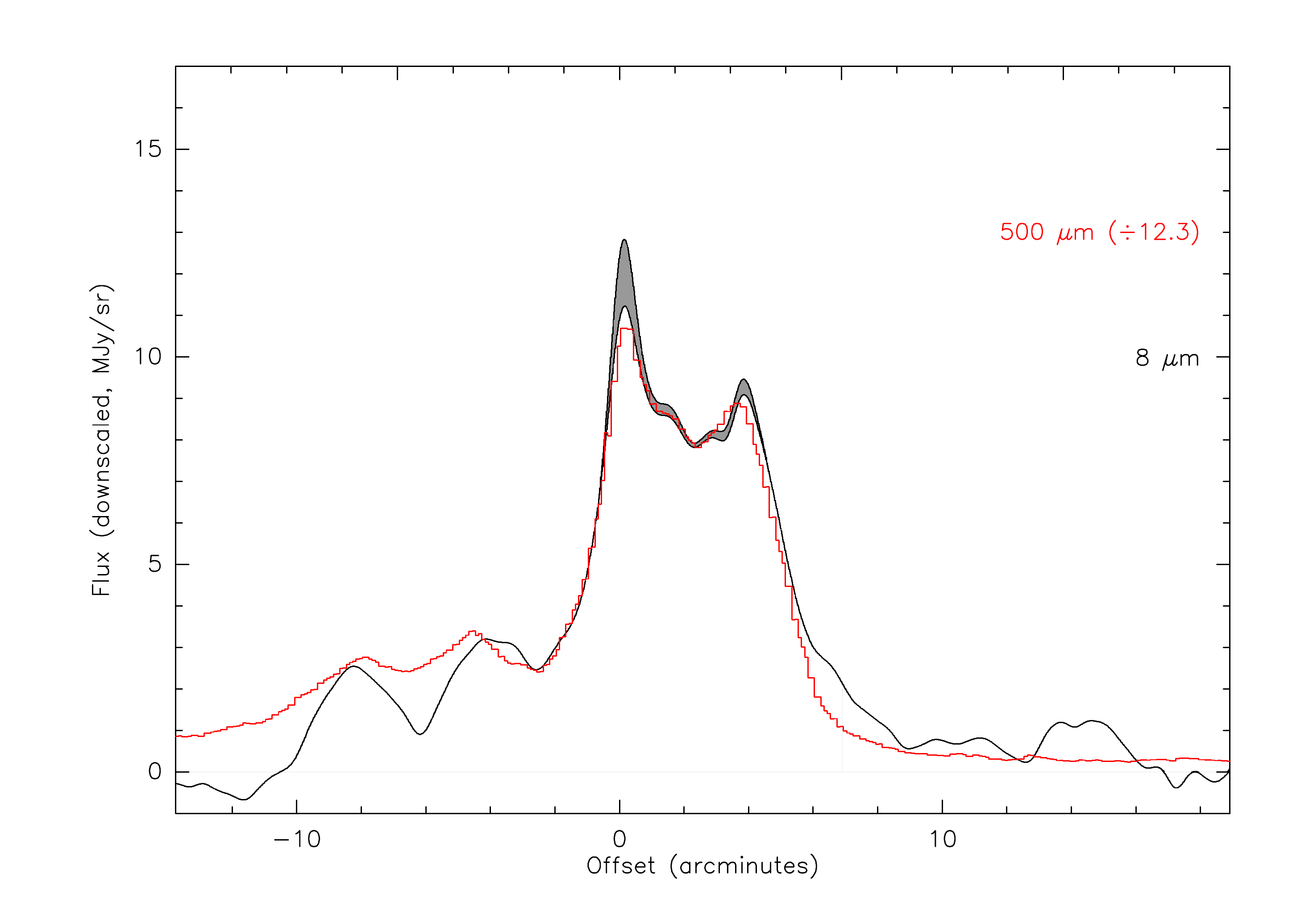}%
   \end{ocg}%
   \hspace{-0.8\hsize}%
 \begin{ocg}{fig:LABOCA}{fig:LABOCA}{0}%
    \includegraphics[width=0.8\hsize]{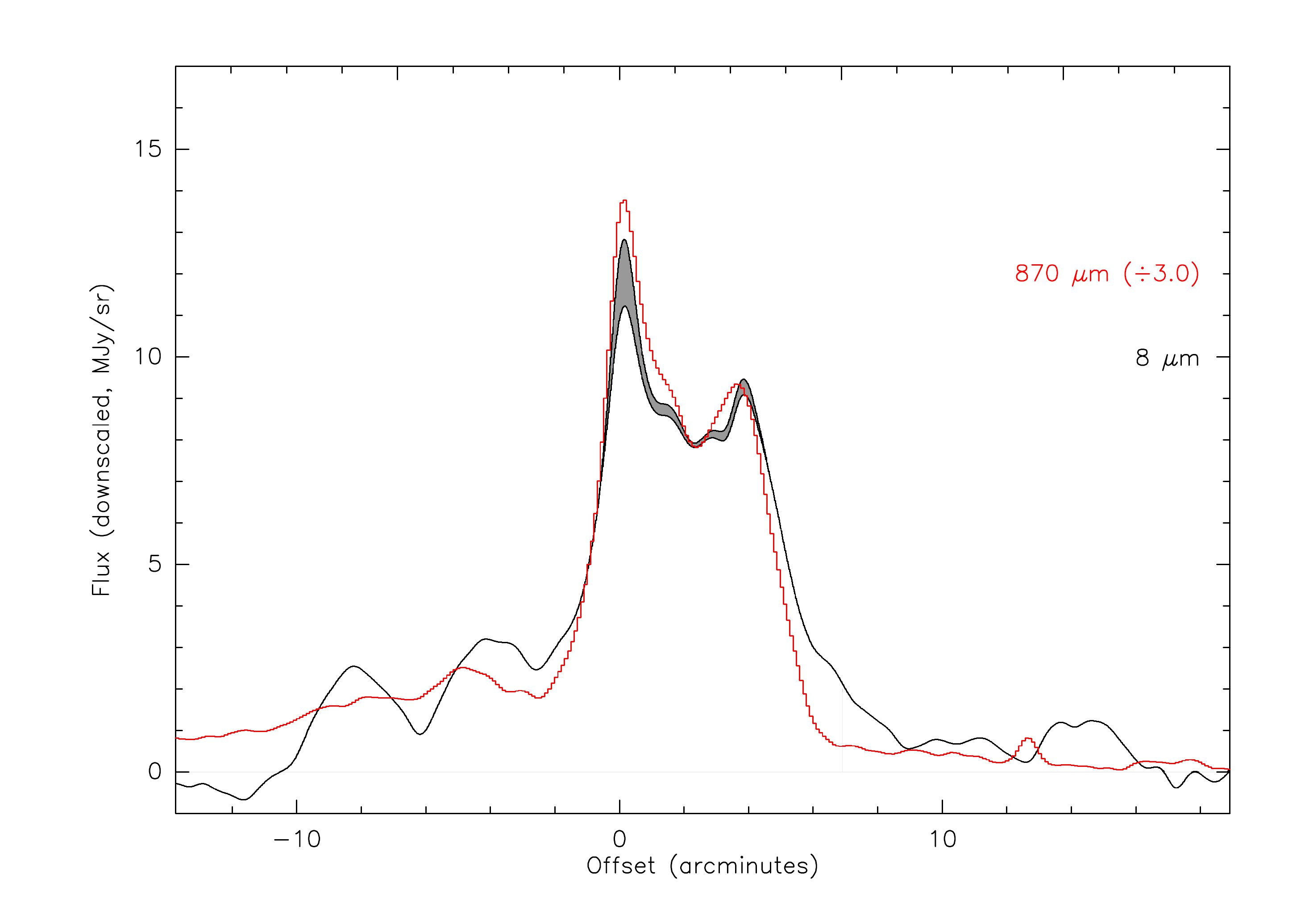}%
  \end{ocg}%
  \hspace{-0.8\hsize}%
 \begin{ocg}{fig:MAMBO}{fig:MAMBO}{0}%
    \includegraphics[width=0.8\hsize]{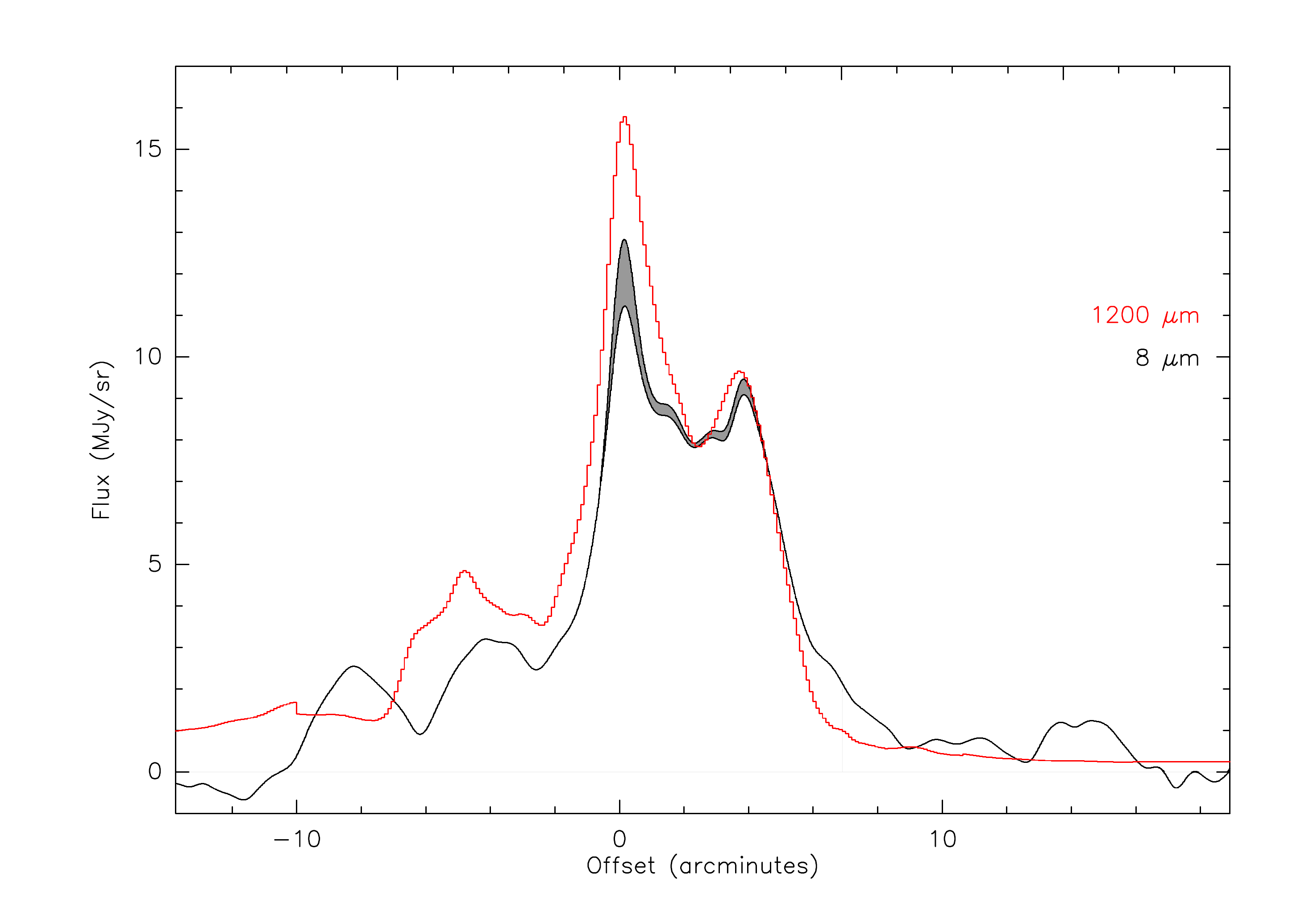}%
  \end{ocg}%
  \hspace{-0.8\hsize}%
   \begin{ocg}{fig:PACSg}{fig:PACSg}{1}%
      \includegraphics[width=0.8\hsize]{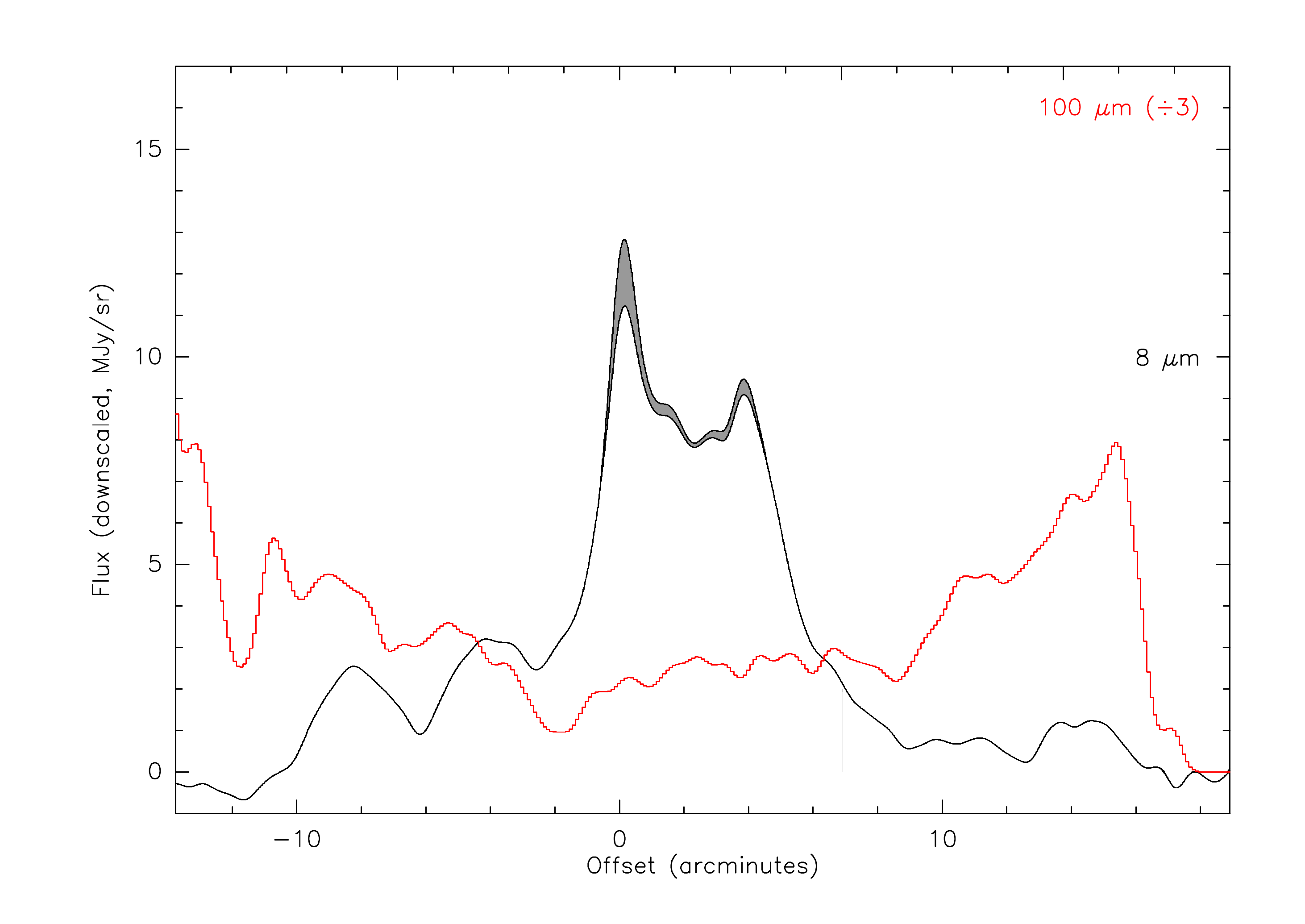}%
    \end{ocg}%
  \caption{L183 cut for dust emission. This cut includes data from \textit{Herschel} PACS and SPIRE, APEX/LABOCA, and IRAM-30m/MAMBO data, all convolved to 37\arcsec\ resolution, shown one by one. The fluxes are scaled down by the amount indicated to the right of the wavelengths to align all the fluxes on the saddle point (except at 100 \mic). The 8 \mic\ opacity range is displayed (black lines filled in grey), the 250 \mic\ is traced in blue for easier comparison with Fig. \ref{fig:L183_8_250_870um_3y}.  \ToggleLayer{fig:PACSr,fig:SPIRES,fig:SPIREM,fig:SPIREL,fig:LABOCA,fig:MAMBO,fig:PACSg}{\protect\cdbox{\textsc{Click here to change wavelength}}}. The FIR wavelengths (except the 100 $\mu$m one) are shown together in the Appendix (Fig. \ref{fig:App_L183_Emission})}
\label{fig:L183_Emission}
\vspace*{-0.3cm}
\end{figure}

\begin{figure}[tp!]
  \centering
  \begin{ocg}{fig:L183_2_SEDs_N_PSC}{fig:L183_2_SEDs_N_PSC}{0}%
    \includegraphics[width=0.7\hsize]{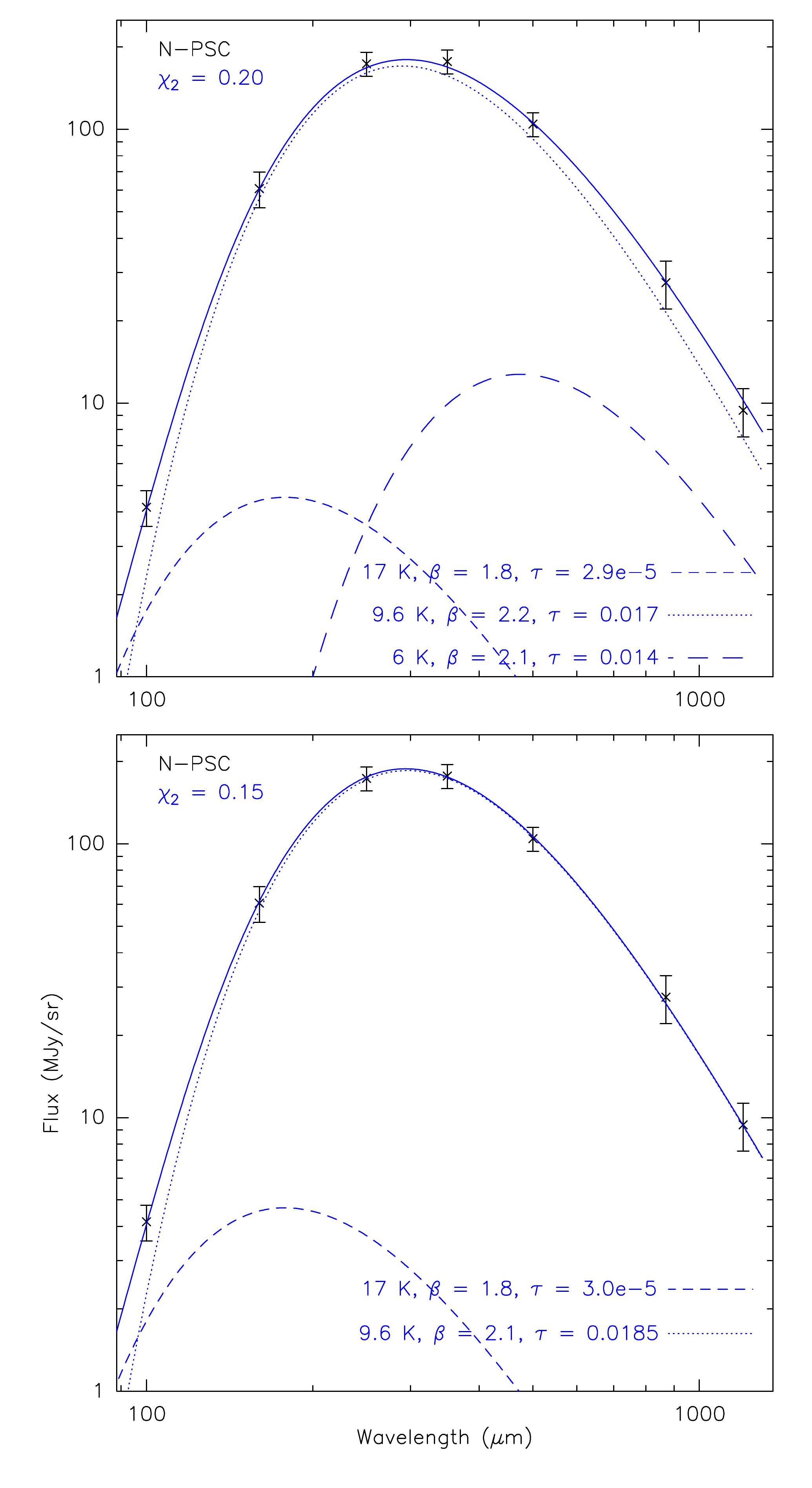}%
  \end{ocg}%
  \hspace{-0.7\hsize}%
   \begin{ocg}{fig:L183_2_SEDs_saddle}{fig:L183_2_SEDs_saddle}{0}%
    \includegraphics[width=0.7\hsize]{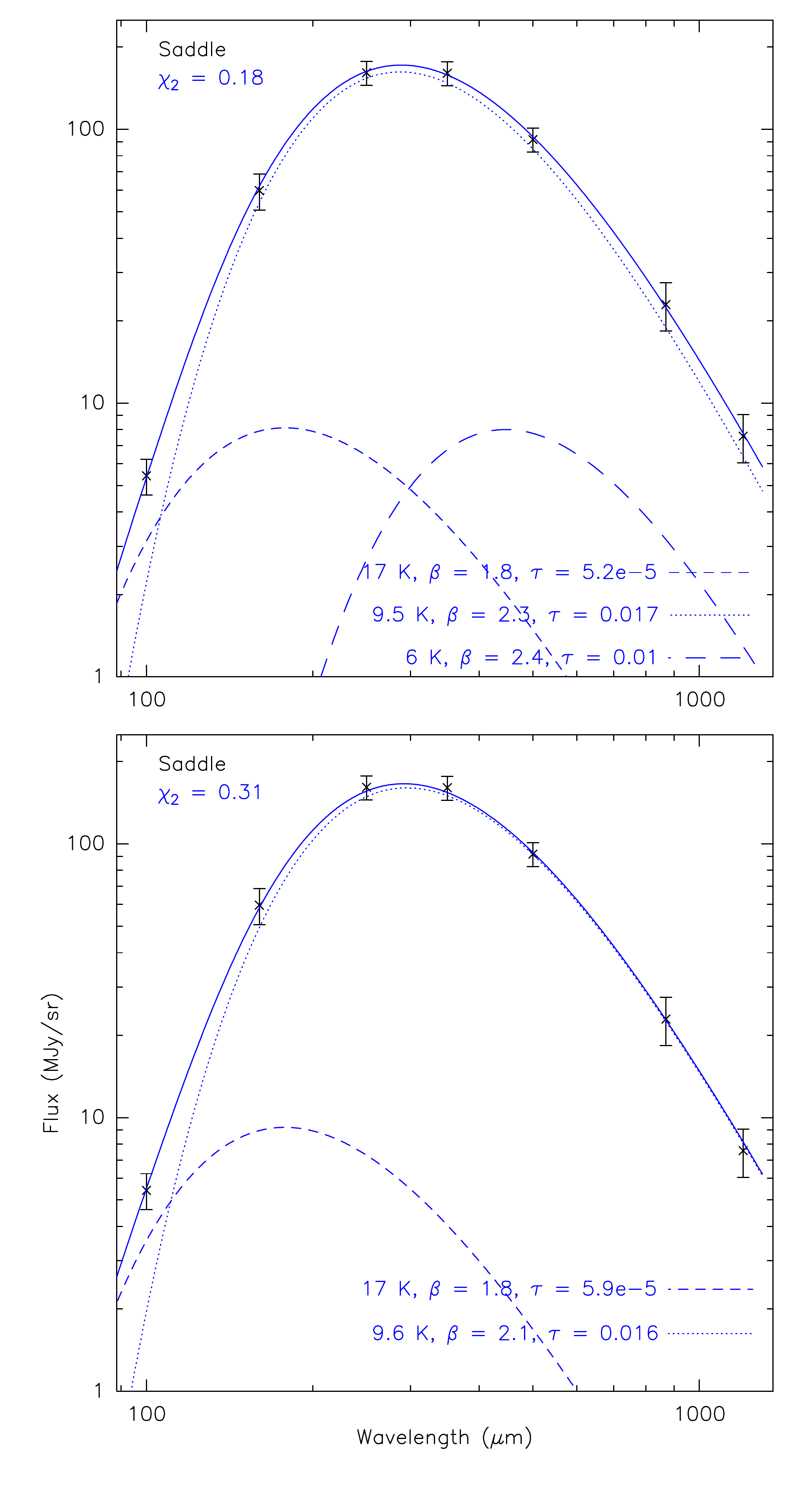}%
  \end{ocg}%
  \hspace{-0.7\hsize}%
   \begin{ocg}{fig:L183_2_SEDs_PSC}{fig:L183_2_SEDs_PSC}{1}%
    \includegraphics[width=0.7\hsize]{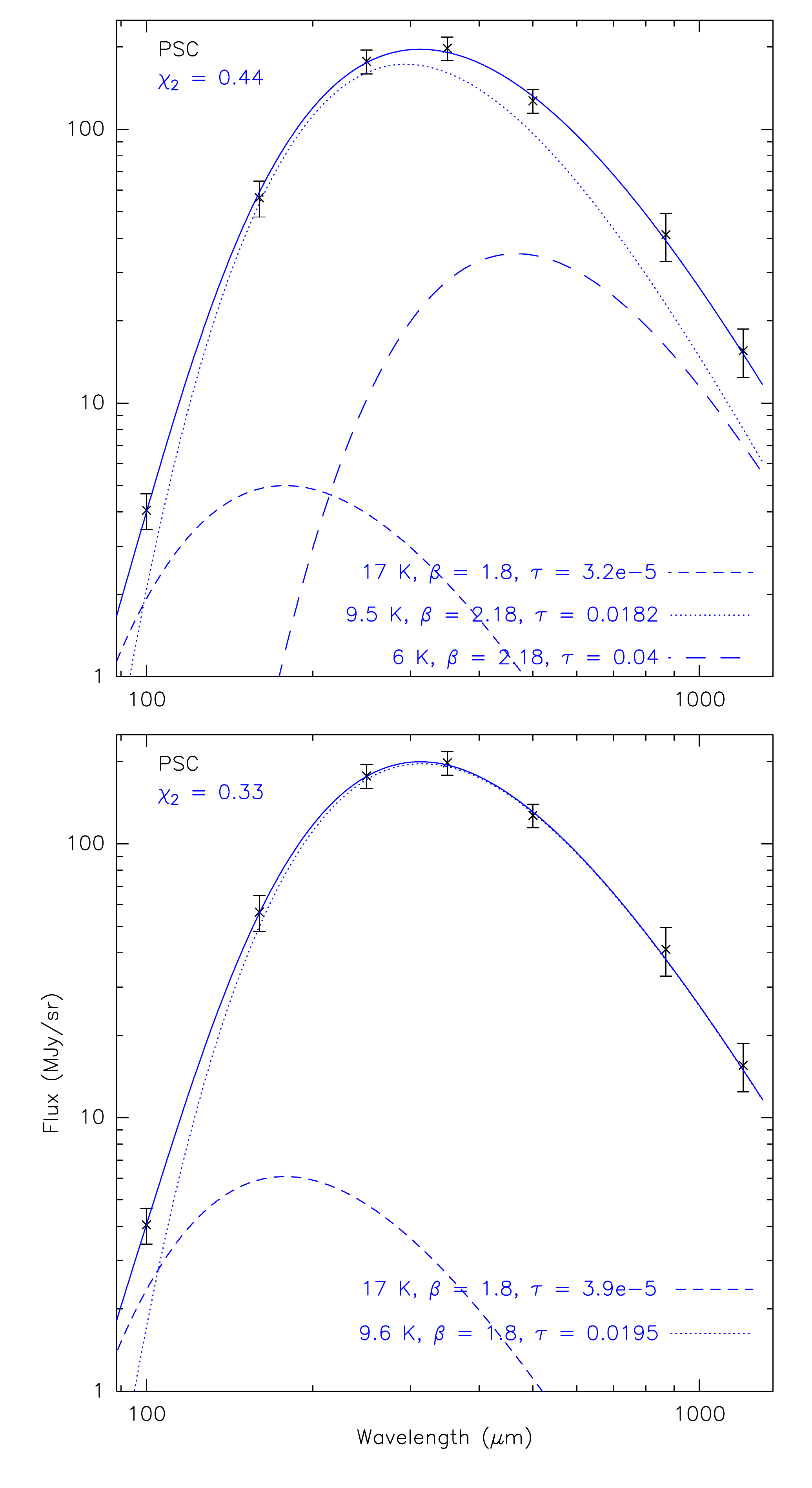}%
  \end{ocg}%
  \caption{SEDs of the three points of interest defined in Fig. \ref{fig:L183_8_250_870um_3y} along the cut (in apparition order, the main PSC, the saddle and the northern PSC). Top row, the SEDs are fitted with three modified blackbodies, bottom row, with two modified blackbodies. Opacity is given at 300 \mic.
    \ToggleLayer{fig:L183_2_SEDs_N_PSC,fig:L183_2_SEDs_saddle,fig:L183_2_SEDs_PSC}{\protect\cdbox{\textsc{Click to display the three cases in turn}}}. The three plots are shown separately in the Appendix (Fig. \ref{fig:App_L183_6_SEDs}).}
  \label{fig:L183_6_SEDs}
\vspace*{-0.4cm}
\end{figure}

While the strong features from the 8 \mic\ data and the 870 \mic\ data are clearly correlated, the 
weak peaks southward (negative offsets in Fig. \ref{fig:L183_8_250_870um_3y}) seen at 8 \mic\ are 
missing at 870 \mic. This is due to the loss of extended features with LABOCA. These features are too 
small to be resolved by \textit{Planck} and are detected only as a smooth increase in emission in this \textit{Planck}+LABOCA combined data cut. The 250 \mic\ 
emission detects the two southern peaks (with a small shift for the most opaque one, most probably due to 
an anisotropic heating). However, it does not correctly trace  the main PSC since the main peak and the northern 
peak have almost the same intensity instead of the ratio 1.4--1.7 we expect from submm observations. Both peaks are also too weak compared to the saddle intensity.  This is reminiscent of the 
similar result reported by \cite{Pagani2004} in which they showed that the ISOPHOT 200 \mic\ map 
does not follow the dust column density in the inner part of the cloud. The low resolution (90\arcsec) of the ISOPHOT map
could have been a partial explanation, but it can now be seen that resolution is not an issue and the 
dust is simply too cold to contribute significantly to the 250 \mic\ emission in the central parts of the 
cloud. Indeed, the peak intensity ratio between the southern peak at offset -4\arcmin40\arcsec\ and the two 
main peaks is about 2 at 250 \mic, while it is larger than 3 for the northern peak and in the range 4--5 
for the main PSC in terms of relative opacity at 8 \mic.
Figure \ref{fig:L183_Emission} shows the dust emission along the cut at wavelengths from 100 to 
1200 \mic. The resolution of all the data has been aligned on that of the SPIRE 500 \mic\ channel 
(37\arcsec). 
The longer the wavelength, the better the tracing of the two dust peaks.
Figure 
\ref{fig:L183_6_SEDs} displays the SEDs for the three points of interest along the cut: the main PSC, 
the saddle, and the northern PSC. For all three positions, we tried to fit the SED with {either two or three modified blackbodies, one at 17\,K, one at $\sim$10\,K, and an optional one at 6\,K,}
\begin{equation}
I_\nu = \sum_{i=1}^n B_\nu(T_{d,i})\tau_{\nu_0}(\frac{\nu}{\nu_0})^{\beta_i} = \sum_{i=1}^n B_\nu(T_{d,i})\kappa_{\nu_0}\mu m_HN(H_{2,i}) (\frac{\nu}{\nu_0})^{\beta_i},
\end{equation}
{with  n = 2 or 3. The parameter B$_\nu$ is the Planck function at dust temperature T$_d$,  $\nu_0$\,=\,1\,THz $ (\lambda$ = 300 \mic), $\mu$ = 2.33   
the mean molecular weight, m$_H$ = 1.67\,\pdix{-24}\,g  the proton weight, N(H$_2$) the gas column density, 
and $\kappa_{\nu_0}$ the dust opacity at 300 \mic.  The 
dust opacity $\kappa_{\nu_0}$ = 0.111 cm$^2$\,g$^{-1}$   corresponds to the thin ice case at density 1\,\pdix{6}\ccb from 
\citet{Ossenkopf1994} for a gas-to-dust ratio of 133 \citep{Compiegne:2011jf}, which is compatible with 
\citet{Ysard2013}. The spectral index $\beta$ describes the modification of the dust opacity $\kappa_\nu$ with frequency. The main blackbody is optimized for all 
three parameters, T$_{d}, \beta$, N(H$_2$).  The cold blackbody, if included, is set at 6\,K, and $\beta$ and 
N(H$_2$) can be adjusted.  The third blackbody at 17\,K, with $\beta = 1.8$, is set to fit the  grains on the cloud surface ,which contribute at 100 \mic.  Their only free 
parameter is N(H$_2$). Their temperature and spectral index are typical of diffuse and cloud surface dust temperatures \citep{Zucconi2001,PlanckCollaboration2011XXIV}. Their contribution is always less than 0.3\% of the total mass and is not discussed any further. We independently adjust 
$\beta$ values in the range 1.5 -- 4 with the constraint that 
$\beta_{T_d} \leq \beta_{6 K}$}. The fits are  optimized so that $\chi^2 \leq 0.5$ in all cases ({all 
parameter values }are in Fig. \ref{fig:L183_6_SEDs}). 
The striking result is that the fit is just as good {with n = 2 (blackbodies at $\sim$\,10 and 17\,K) or n = 3 
(6, $\sim$\,10, and 17\,K). The $\chi^2$ values remain basically identical.  Towards the main PSC, the opacity at 
300 
\mic\ varies from 0.02 $\pm$0.07 (n = 2) to 
0.058 (n = 3), almost a factor of three, in the case presented in Fig. \ref{fig:L183_6_SEDs}, and could go up to 0.1 (a factor of five higher) in the extreme case where $\beta_{6 K}$ $\approx$ 3.9.} As expected the difference is less for the saddle and the northern PSC but 
the increase in opacity can still be $\geq$\,50\%. There is no way to 
discriminate between the fits simply based on these data. 
 Of course, three blackbody fitting instead of two is still far from  the real case (a continuous variation of temperature from $\sim$\,17\,K at the surface of the clouds to $\sim$\,6\,K in the densest cores, \citealt{Zucconi2001,Evans2001}) but comes much closer to it already.
 
{The fits we find with an imposed 6\,K blackbody are not unique. Since a solution can be found without a 6 K component, the opacity of this component can be varied from 0 up to 0.1 depending on the other parameters, T$_{dust}$ and $\beta$ in particular, with similar $\chi^2$ values.  Constraints must come from other observations, which we discuss now.}

 {For the main PSC, N(H$_2$) is in the 
range 4.5\,$\pm$1.7\,\pdixb{22}\,\sqcb for n = 2 to 1.3$_{-0.8}^{+1.0}$\,\pdixb{23} \sqcb for n = 3. The  second result encompasses the column density }
determined via gas modelling. \citet{Pagani2007}  indicate N(H$_2$)  
$\sim$1.0\,\pdixb{23}\,\sqcb from \ndhpb data when averaged in a 37\arcsec\ beam (after correcting for the new \ndhp--H$_2$ collisional coefficients, \citealt{Lique2015}). However  \ndhpb only traces the dense region 
and not the envelope of the cloud. From 
\cdhob measurements \citep{Pagani2005}, we infer a cloud envelope column density of 
$\sim$2\,\pdix{22}\,\sqc. The total is $\sim$1.2\,\pdix{23}\,\sqc. The difference with the  
n = 2 fit (ratio of 2.5) is larger than the different uncertainties involved here. With the new \ndhp--H$_2$ collisional coefficients, we also found that the gas 
temperature in the core is 6\,K (7\,K in 
\citealt{Pagani2007}) where the density is $\geq$ 5\,\pdix 5\,\cc, {which is a density high enough to efficiently thermalise gas and dust (n $>$ 1\pdix{5}\,\cc\ is required, \citealt{Goldsmith:2001gh})}. Therefore, dust and gas temperatures and column densities {can be made }consistent only if 
we introduce a 6\,K dust component in the fit.
Based only on \textit{Herschel} data, \citet{Roy:2014bv} also find a temperature lower than that given 
by a single SED fit (9.8\,K instead of 11.6\,K in the case of L1689B at its centre), and their fit is more 
realistic than the one we present here by using a continuously varying temperature with radius. 
However, despite this more sophisticated analysis,  Roy et al.'s method fails to reveal the very cold dust 
(6--8\,K) in the heart of L1689B that we have identified from \ndhpb measurements (Bacmann et 
al., in prep). These measurements indicate a 37\arcsec\ beam--averaged column density of 7--11\,\pdixb{22}\,\sqc, again a factor 2 to 3 times higher than Roy et al.'s 3.5\,\pdixb{22}\,\sqc, which does not even take the contribution from the envelope devoid of \ndhpb emission into account. {Another attempt by \citet{Marsh:2014dc} gives better results. They trace cold dust down to 6 K but they use a priori density and temperature profiles and therefore their results somewhat depend on their input parameters  in contrast with  Roy et al. approach.}

It is  clear from these two cases that very cold dust exists and is not identified as such solely by its  emission.
This is because of the well-known fact that warmer dust {(even as low as 10 K)} outshines the very cold dust because of the 
non-linearity of the blackbody function (at 6 K, the Rayleigh-Jeans approximation is not valid above 
100 GHz). This is clearly visible in Fig. \ref{fig:L183_6_SEDs} where the 6\,K blackbody contribution
is smaller than the  $\sim$\,10\,K contribution,  even at $\lambda$ = 1.2 mm and even when it represents 70\% of the total mass (45\% of the signal in that case). Though L183 and L1689B could be 
thought of as peculiar cases, we believe they are only clear illustrations of a general problem, since dust below 9 
K is predicted to occur for radial extinctions as low as 
\Av = 10 mag \citep{Zucconi2001}, in standard ISRF conditions (no local heating source). It is 
clear that studies of filaments, and cold dark clouds in general, based solely on dust emission (and worse, single 
component SED fits), even including \textit{Herschel} or \textit{Planck} data, should meet the 
 degeneracy problem we 
expose here and will miss a large fraction of the mass in these objects. Even more sophisticated 
modelling, as presented by \citet{Roy:2014bv}, misses a large fraction of the mass. In the present 
two cases, L183 and L1689B, 30 to 70\% of the total 
mass is lacking. The dust column density must be retrieved by other means (dust in extinction, 
molecular 
 emission). Therefore, critical and 
 threshold masses, core and filament stabilities,  and density profiles in studies limited to dust 
 emission  should be considered with caution.

 \begin{acknowledgements}
      CL acknowledges financial support by the DIM ACAV and \textit{R\'egion \^Ile de France}. MJ and and V-MP acknowledge the support of Academy of Finland grant 250741. We thank the referee for her/his remarks which helped to clarify this Letter.
 \end{acknowledgements}
\vspace{-0.4cm}
 \bibliographystyle{aa}
 
 \bibliography{/Users/laurent/bibtex/references,/Users/laurent/bibtex/papers}

\begin{thebibliography}{50}
\expandafter\ifx\csname natexlab\endcsname\relax\def\natexlab#1{#1}\fi

\bibitem[{{Andre} {et~al.}(1993){Andre}, {Ward-Thompson}, \&
  {Barsony}}]{Andre1993}
{Andre}, P., {Ward-Thompson}, D., \& {Barsony}, M. 1993, \apj, 406, 122

\bibitem[{{Bacmann} {et~al.}(2000){Bacmann}, {Andr{\'e}}, {Puget}, {Abergel},
  {Bontemps}, \& {Ward-Thompson}}]{Bacmann2000}
{Bacmann}, A., {Andr{\'e}}, P., {Puget}, J.-L., {et~al.} 2000, \aap, 361, 555

\bibitem[{{Bok}(1956)}]{Bok1956}
{Bok}, B.~J. 1956, \aj, 61, 309

\bibitem[{{Bok} \& {Cordwell}(1973)}]{Bok1973}
{Bok}, B.~J. \& {Cordwell}, C.~S. 1973, in Molecules in the Galactic
  Environment, ed. M.~A. {Gordon} \& L.~E. {Snyder}, 54--+

\bibitem[{Brady~Ford \& Shirley(2011)}]{BradyFord:2011ke}
Brady~Ford, A. \& Shirley, Y.~L. 2011, ApJ, 728, 144

\bibitem[{Compi{\`e}gne {et~al.}(2011)Compi{\`e}gne, Flagey, Verstraete, Jones,
  Bernard, Boulanger, Le~Bourlot, Paradis, \& Ysard}]{Compiegne:2011jf}
Compi{\`e}gne, M., Flagey, N., Verstraete, L., {et~al.} 2011, A{\&}A, 525, A103

\bibitem[{{Evans} {et~al.}(2001){Evans}, {Rawlings}, {Shirley}, \&
  {Mundy}}]{Evans2001}
{Evans}, II, N.~J., {Rawlings}, J.~M.~C., {Shirley}, Y.~L., \& {Mundy}, L.~G.
  2001, \apj, 557, 193

\bibitem[{Fazio {et~al.}(2004)Fazio, Hora, Allen, Ashby, Barmby, Deutsch,
  Huang, Kleiner, Marengo, Megeath, Melnick, Pahre, Patten, Polizotti, Smith,
  Taylor, Wang, Willner, Hoffmann, Pipher, Forrest, McMurty, McCreight,
  McKelvey, McMurray, Koch, Moseley, Arendt, Mentzell, Marx, Losch, Mayman,
  Eichhorn, Krebs, Jhabvala, Gezari, Fixsen, Flores, Shakoorzadeh, Jungo,
  Hakun, Workman, Karpati, Kichak, Whitley, Mann, Tollestrup, Eisenhardt,
  Stern, Gorjian, Bhattacharya, Carey, Nelson, Glaccum, Lacy, Lowrance, Laine,
  Reach, Stauffer, Surace, Wilson, Wright, Hoffman, Domingo, \&
  Cohen}]{Fazio:2004eb}
Fazio, G.~G., Hora, J.~L., Allen, L.~E., {et~al.} 2004, ApJS, 154, 10

\bibitem[{{Flower} {et~al.}(2005){Flower}, {Pineau Des For{\^e}ts}, \&
  {Walmsley}}]{Flower2005}
{Flower}, D.~R., {Pineau Des For{\^e}ts}, G., \& {Walmsley}, C.~M. 2005, \aap,
  436, 933

\bibitem[{Goldsmith(2001)}]{Goldsmith:2001gh}
Goldsmith, P.~F. 2001, ApJ, 557, 736

\bibitem[{{Griffin} {et~al.}(2010){Griffin}, {Abergel}, {Abreu}, {Ade},
  {Andr{\'e}}, {Augueres}, {Babbedge}, {Bae}, {Baillie}, {Baluteau}, {Barlow},
  {Bendo}, {Benielli}, {Bock}, {Bonhomme}, {Brisbin}, {Brockley-Blatt},
  {Caldwell}, {Cara}, {Castro-Rodriguez}, {Cerulli}, {Chanial}, {Chen},
  {Clark}, {Clements}, {Clerc}, {Coker}, {Communal}, {Conversi}, {Cox},
  {Crumb}, {Cunningham}, {Daly}, {Davis}, {de Antoni}, {Delderfield}, {Devin},
  {di Giorgio}, {Didschuns}, {Dohlen}, {Donati}, {Dowell}, {Dowell}, {Duband},
  {Dumaye}, {Emery}, {Ferlet}, {Ferrand}, {Fontignie}, {Fox}, {Franceschini},
  {Frerking}, {Fulton}, {Garcia}, {Gastaud}, {Gear}, {Glenn}, {Goizel},
  {Griffin}, {Grundy}, {Guest}, {Guillemet}, {Hargrave}, {Harwit}, {Hastings},
  {Hatziminaoglou}, {Herman}, {Hinde}, {Hristov}, {Huang}, {Imhof}, {Isaak},
  {Israelsson}, {Ivison}, {Jennings}, {Kiernan}, {King}, {Lange}, {Latter},
  {Laurent}, {Laurent}, {Leeks}, {Lellouch}, {Levenson}, {Li}, {Li},
  {Lilienthal}, {Lim}, {Liu}, {Lu}, {Madden}, {Mainetti}, {Marliani}, {McKay},
  {Mercier}, {Molinari}, {Morris}, {Moseley}, {Mulder}, {Mur}, {Naylor},
  {Nguyen}, {O'Halloran}, {Oliver}, {Olofsson}, {Olofsson}, {Orfei}, {Page},
  {Pain}, {Panuzzo}, {Papageorgiou}, {Parks}, {Parr-Burman}, {Pearce},
  {Pearson}, {P{\'e}rez-Fournon}, {Pinsard}, {Pisano}, {Podosek}, {Pohlen},
  {Polehampton}, {Pouliquen}, {Rigopoulou}, {Rizzo}, {Roseboom}, {Roussel},
  {Rowan-Robinson}, {Rownd}, {Saraceno}, {Sauvage}, {Savage}, {Savini},
  {Sawyer}, {Scharmberg}, {Schmitt}, {Schneider}, {Schulz}, {Schwartz},
  {Shafer}, {Shupe}, {Sibthorpe}, {Sidher}, {Smith}, {Smith}, {Smith},
  {Spencer}, {Stobie}, {Sudiwala}, {Sukhatme}, {Surace}, {Stevens}, {Swinyard},
  {Trichas}, {Tourette}, {Triou}, {Tseng}, {Tucker}, {Turner}, {Vaccari},
  {Valtchanov}, {Vigroux}, {Virique}, {Voellmer}, {Walker}, {Ward}, {Waskett},
  {Weilert}, {Wesson}, {White}, {Whitehouse}, {Wilson}, {Winter}, {Woodcraft},
  {Wright}, {Xu}, {Zavagno}, {Zemcov}, {Zhang}, \& {Zonca}}]{Griffin2010}
{Griffin}, M.~J., {Abergel}, A., {Abreu}, A., {et~al.} 2010, \aap, 518, L3+

\bibitem[{Juvela {et~al.}(2006)Juvela, Pelkonen, Padoan, \&
  Mattila}]{Juvela2006}
Juvela, M., Pelkonen, V.-M., Padoan, P., \& Mattila, K. 2006, A{\&}A, 457, 877

\bibitem[{{Juvela} {et~al.}(2010){Juvela}, {Ristorcelli}, {Montier},
  {Marshall}, {Pelkonen}, {Malinen}, {Ysard}, {T{\'o}th}, {Harju}, {Bernard},
  {Schneider}, {Vereb{\'e}lyi}, {Anderson}, {Andr{\'e}}, {Giard}, {Krause},
  {Lehtinen}, {Macias-Perez}, {Martin}, {McGehee}, {Meny}, {Motte}, {Pagani},
  {Paladini}, {Reach}, {Valenziano}, {Ward-Thompson}, \&
  {Zavagno}}]{Juvela2010}
{Juvela}, M., {Ristorcelli}, I., {Montier}, L.~A., {et~al.} 2010, \aap, 518,
  L93+

\bibitem[{Juvela {et~al.}(2012)Juvela, Ristorcelli, Pagani, Doi, Pelkonen,
  Marshall, Bernard, Falgarone, Malinen, Marton, McGehee, Montier, Motte,
  Paladini, T{\'o}th, Ysard, Zahorecz, \& Zavagno}]{2012A&A...541A..12J}
Juvela, M., Ristorcelli, I., Pagani, L., {et~al.} 2012, A{\&}A, 541, 12

\bibitem[{Juvela \& Ysard(2012)}]{Juvela:2012ex}
Juvela, M. \& Ysard, N. 2012, A{\&}A, 541, A33

\bibitem[{Juvela {et~al.}(2014)}]{Juvela-subm}
Juvela, M., Demyk, K., Doy, Y., {et~al.} 2014, A\&A, subm. 

\bibitem[{{Lada}(1987)}]{Lada1987}
{Lada}, C.~J. 1987, in IAU Symposium, Vol. 115, Star Forming Regions, ed.
  M.~{Peimbert} \& J.~{Jugaku}, 1--17

\bibitem[{{Lada} {et~al.}(1994){Lada}, {Lada}, {Clemens}, \&
  {Bally}}]{Lada1994}
{Lada}, C.~J., {Lada}, E.~A., {Clemens}, D.~P., \& {Bally}, J. 1994, \apj, 429,
  694

\bibitem[{{Lamarre} {et~al.}(2010){Lamarre}, {Puget}, {Ade}, {Bouchet},
  {Guyot}, {Lange}, {Pajot}, {Arondel}, {Benabed}, {Beney}, {Beno{\^i}t},
  {Bernard}, {Bhatia}, {Blanc}, {Bock}, {Br{\'e}elle}, {Bradshaw}, {Camus},
  {Catalano}, {Charra}, {Charra}, {Church}, {Couchot}, {Coulais}, {Crill},
  {Crook}, {Dassas}, {de Bernardis}, {Delabrouille}, {de Marcillac}, {Delouis},
  {D{\'e}sert}, {Dumesnil}, {Dupac}, {Efstathiou}, {Eng}, {Evesque},
  {Fourmond}, {Ganga}, {Giard}, {Gispert}, {Guglielmi}, {Haissinski},
  {Henrot-Versill{\'e}}, {Hivon}, {Holmes}, {Jones}, {Koch}, {Lagard{\`e}re},
  {Lami}, {Land{\'e}}, {Leriche}, {Leroy}, {Longval},
  {Mac{\'{\i}}as-P{\'e}rez}, {Maciaszek}, {Maffei}, {Mansoux}, {Marty}, {Masi},
  {Mercier}, {Miville-Desch{\^e}nes}, {Moneti}, {Montier}, {Murphy},
  {Narbonne}, {Nexon}, {Paine}, {Pahn}, {Perdereau}, {Piacentini}, {Piat},
  {Plaszczynski}, {Pointecouteau}, {Pons}, {Ponthieu}, {Prunet}, {Rambaud},
  {Recouvreur}, {Renault}, {Ristorcelli}, {Rosset}, {Santos}, {Savini},
  {Serra}, {Stassi}, {Sudiwala}, {Sygnet}, {Tauber}, {Torre}, {Tristram},
  {Vibert}, {Woodcraft}, {Yurchenko}, \& {Yvon}}]{Lamarre2010}
{Lamarre}, J.-M., {Puget}, J.-L., {Ade}, P.~A.~R., {et~al.} 2010, \aap, 520, A9

\bibitem[{{Lef{\`e}vre} {et~al.}(2014){Lef{\`e}vre}, {Pagani}, {Juvela},
  {Paladini}, {Lallement}, {Marshall}, {Andersen}, {Bacmann}, {McGehee},
  {Montier}, {Noriega-Crespo}, {Pelkonen}, {Ristorcelli}, \&
  {Steinacker}}]{Lef`evre2014}
{Lef{\`e}vre}, C., {Pagani}, L., {Juvela}, M., {et~al.} 2014, ArXiv e-prints

\bibitem[{Lehtinen \& Mattila(1996)}]{Lehtinen:1996ti}
Lehtinen, K. \& Mattila, K. 1996, A{\&}A, 309, 570

\bibitem[{{Lemme} {et~al.}(1995){Lemme}, {Walmsley}, {Wilson}, \&
  {Muders}}]{Lemme1995}
{Lemme}, C., {Walmsley}, C.~M., {Wilson}, T.~L., \& {Muders}, D. 1995, \aap,
  302, 509

\bibitem[{{Lique} {et~al.}(2015){Lique}, {Daniel}, {Pagani}, \&
  {Feautrier}}]{Lique2015}
{Lique}, F., {Daniel}, F., {Pagani}, L., \& {Feautrier}, N. 2015, \mnras, 446,
  1245

\bibitem[{Lombardi(2009)}]{Lombardi:2009hd}
Lombardi, M. 2009, A{\&}A, 493, 735

\bibitem[{Lombardi \& Alves(2001)}]{Lombardi:2001bka}
Lombardi, M. \& Alves, J. 2001, A{\&}A, 377, 1023

\bibitem[{Malinen {et~al.}(2011)Malinen, Juvela, Collins, Lunttila, \&
  Padoan}]{Malinen:2011eu}
Malinen, J., Juvela, M., Collins, D.~C., Lunttila, T., \& Padoan, P. 2011,
  A{\&}A, 1

\bibitem[{Marsh {et~al.}(2014)Marsh, Griffin, Palmeirim, Andre, Kirk,
  Stamatellos, Ward-Thompson, Roy, Bontemps, Francesco, Elia, Hill,
  K{\"o}nyves, Motte, Nguyen~Luong, Peretto, Pezzuto, Rivera-Ingraham,
  Schneider, Spinoglio, \& White}]{Marsh:2014dc}
Marsh, K.~A., Griffin, M.~J., Palmeirim, P., {et~al.} 2014, MNRAS, 439, 3683

\bibitem[{Miville-Desch{\^e}nes \& Lagache(2005)}]{MivilleDeschenes:2005fr}
Miville-Desch{\^e}nes, M.-A. \& Lagache, G. 2005, ApJS, 157, 302

\bibitem[Montillaud {et~al.}(2014)]{Montillaud-subm}Montillaud, J., Juvela, M., Rivera-Ingraham, A., et al., 2014, A\&A, subm.

\bibitem[{{Murakami} {et~al.}(2007){Murakami}, {Baba}, {Barthel}, {Clements},
  {Cohen}, {Doi}, {Enya}, {Figueredo}, {Fujishiro}, {Fujiwara}, {Fujiwara},
  {Garcia-Lario}, {Goto}, {Hasegawa}, {Hibi}, {Hirao}, {Hiromoto}, {Hong},
  {Imai}, {Ishigaki}, {Ishiguro}, {Ishihara}, {Ita}, {Jeong}, {Jeong},
  {Kaneda}, {Kataza}, {Kawada}, {Kawai}, {Kawamura}, {Kessler}, {Kester},
  {Kii}, {Kim}, {Kim}, {Kobayashi}, {Koo}, {Kwon}, {Lee}, {Lorente}, {Makiuti},
  {Matsuhara}, {Matsumoto}, {Matsuo}, {Matsuura}, {M{\"u}ller}, {Murakami},
  {Nagata}, {Nakagawa}, {Naoi}, {Narita}, {Noda}, {Oh}, {Ohnishi}, {Ohyama},
  {Okada}, {Okuda}, {Oliver}, {Onaka}, {Ootsubo}, {Oyabu}, {Pak}, {Park},
  {Pearson}, {Rowan-Robinson}, {Saito}, {Sakon}, {Salama}, {Sato}, {Savage},
  {Serjeant}, {Shibai}, {Shirahata}, {Sohn}, {Suzuki}, {Takagi}, {Takahashi},
  {Tanab{\'e}}, {Takeuchi}, {Takita}, {Thomson}, {Uemizu}, {Ueno}, {Usui},
  {Verdugo}, {Wada}, {Wang}, {Watabe}, {Watarai}, {White}, {Yamamura},
  {Yamauchi}, \& {Yasuda}}]{Murakami2007}
{Murakami}, H., {Baba}, H., {Barthel}, P., {et~al.} 2007, \pasj, 59, 369

\bibitem[{Nielbock {et~al.}(2012)Nielbock, Launhardt, Steinacker, Lippok,
  Stutz, Balog, Beuther, Bouwman, Henning, Hily-Blant, Kainulainen, Krause,
  Linz, Ragan, Risacher, \& Schmiedeke}]{Nielbock:2012ih}
Nielbock, M., Launhardt, R., Steinacker, J., {et~al.} 2012, A{\&}A, 547, A11

\bibitem[{{Ossenkopf} \& {Henning}(1994)}]{Ossenkopf1994}
{Ossenkopf}, V. \& {Henning}, T. 1994, \aap, 291, 943

\bibitem[{{Ott}(2010)}]{Ott2010}
{Ott}, S. 2010, in Astronomical Society of the Pacific Conference Series, Vol.
  434, Astronomical Data Analysis Software and Systems XIX, ed. Y.~{Mizumoto},
  K.-I. {Morita}, \& M.~{Ohishi}, 139

\bibitem[{{Pagani} {et~al.}(2007){Pagani}, {Bacmann}, {Cabrit}, \&
  {Vastel}}]{Pagani2007}
{Pagani}, L., {Bacmann}, A., {Cabrit}, S., \& {Vastel}, C. 2007, \aap, 467, 179

\bibitem[{{Pagani} {et~al.}(2004){Pagani}, {Bacmann}, {Motte}, {Cambr{\'e}sy},
  {Fich}, {Lagache}, {Miville-Desch{\^e}nes}, {Pardo}, \&
  {Apponi}}]{Pagani2004}
{Pagani}, L., {Bacmann}, A., {Motte}, F., {et~al.} 2004, \aap, 417, 605

\bibitem[{{Pagani} {et~al.}(2005){Pagani}, {Pardo}, {Apponi}, {Bacmann}, \&
  {Cabrit}}]{Pagani2005}
{Pagani}, L., {Pardo}, J.-R., {Apponi}, A.~J., {Bacmann}, A., \& {Cabrit}, S.
  2005, \aap, 429, 181

\bibitem[{{Pagani} {et~al.}(2010){Pagani}, {Steinacker}, {Bacmann}, {Stutz}, \&
  {Henning}}]{Pagani2010a}
{Pagani}, L., {Steinacker}, J., {Bacmann}, A., {Stutz}, A., \& {Henning}, T.
  2010, Science, 329, 1622

\bibitem[{{Pilbratt} {et~al.}(2010){Pilbratt}, {Riedinger}, {Passvogel},
  {Crone}, {Doyle}, {Gageur}, {Heras}, {Jewell}, {Metcalfe}, {Ott}, \&
  {Schmidt}}]{Pilbratt2010}
{Pilbratt}, G.~L., {Riedinger}, J.~R., {Passvogel}, T., {et~al.} 2010, \aap,
  518, L1+

\bibitem[{{Planck Collaboration}(2011)}]{PlanckCollaboration2011XXIV}
{Planck Collaboration}. 2011, \aap, 536, A24

\bibitem[{{Poglitsch} {et~al.}(2010){Poglitsch}, {Waelkens}, {Geis},
  {Feuchtgruber}, {Vandenbussche}, {Rodriguez}, {Krause}, {Renotte}, {van
  Hoof}, {Saraceno}, {Cepa}, {Kerschbaum}, {Agn{\`e}se}, {Ali}, {Altieri},
  {Andreani}, {Augueres}, {Balog}, {Barl}, {Bauer}, {Belbachir}, {Benedettini},
  {Billot}, {Boulade}, {Bischof}, {Blommaert}, {Callut}, {Cara}, {Cerulli},
  {Cesarsky}, {Contursi}, {Creten}, {De Meester}, {Doublier}, {Doumayrou},
  {Duband}, {Exter}, {Genzel}, {Gillis}, {Gr{\"o}zinger}, {Henning},
  {Herreros}, {Huygen}, {Inguscio}, {Jakob}, {Jamar}, {Jean}, {de Jong},
  {Katterloher}, {Kiss}, {Klaas}, {Lemke}, {Lutz}, {Madden}, {Marquet},
  {Martignac}, {Mazy}, {Merken}, {Montfort}, {Morbidelli}, {M{\"u}ller},
  {Nielbock}, {Okumura}, {Orfei}, {Ottensamer}, {Pezzuto}, {Popesso},
  {Putzeys}, {Regibo}, {Reveret}, {Royer}, {Sauvage}, {Schreiber}, {Stegmaier},
  {Schmitt}, {Schubert}, {Sturm}, {Thiel}, {Tofani}, {Vavrek}, {Wetzstein},
  {Wieprecht}, \& {Wiezorrek}}]{Poglitsch2010}
{Poglitsch}, A., {Waelkens}, C., {Geis}, N., {et~al.} 2010, \aap, 518, L2+

\bibitem[{{Roussel}(2013)}]{Roussel2013}
{Roussel}, H. 2013, \pasp, 125, 1126

\bibitem[{Roy {et~al.}(2014)Roy, Andr{\'e}, Palmeirim, Attard, K{\"o}nyves,
  Schneider, Peretto, Men{\textquoteright}shchikov, Ward-Thompson, Kirk,
  Griffin, Marsh, Abergel, Arzoumanian, Benedettini, Hill, Motte, Nguyen~Luong,
  Pezzuto, Rivera-Ingraham, Roussel, Rygl, Spinoglio, Stamatellos, \&
  White}]{Roy:2014bv}
Roy, A., Andr{\'e}, P., Palmeirim, P., {et~al.} 2014, A{\&}A, 562, A138

\bibitem[{{Steinacker} {et~al.}(2010){Steinacker}, {Pagani}, {Bacmann}, \&
  {Guieu}}]{Steinacker2010}
{Steinacker}, J., {Pagani}, L., {Bacmann}, A., \& {Guieu}, S. 2010, \aap, 511,
  A9+

\bibitem[{{Stepnik} {et~al.}(2003){Stepnik}, {Abergel}, {Bernard}, {Boulanger},
  {Cambr{\'e}sy}, {Giard}, {Jones}, {Lagache}, {Lamarre}, {Meny}, {Pajot}, {Le
  Peintre}, {Ristorcelli}, {Serra}, \& {Torre}}]{Stepnik2003}
{Stepnik}, B., {Abergel}, A., {Bernard}, J.-P., {et~al.} 2003, \aap, 398, 551

\bibitem[{{Tafalla} {et~al.}(2002){Tafalla}, {Myers}, {Caselli}, {Walmsley}, \&
  {Comito}}]{Tafalla2002}
{Tafalla}, M., {Myers}, P.~C., {Caselli}, P., {Walmsley}, C.~M., \& {Comito},
  C. 2002, \apj, 569, 815

\bibitem[{{Tauber} {et~al.}(2010){Tauber}, {Mandolesi}, {Puget}, {Banos},
  {Bersanelli}, {Bouchet}, {Butler}, {Charra}, {Crone}, {Dodsworth}, \&
  et~al.}]{Tauber2010}
{Tauber}, J.~A., {Mandolesi}, N., {Puget}, J.-L., {et~al.} 2010, \aap, 520, A1+

\bibitem[{{Ward-Thompson} {et~al.}(1994){Ward-Thompson}, {Scott}, {Hills}, \&
  {Andr\'e}}]{Ward-Thompson1994}
{Ward-Thompson}, D., {Scott}, P.~F., {Hills}, R.~E., \& {Andr\'e}, P. 1994,
  \mnras, 268, 276

\bibitem[{{Werner} {et~al.}(2004){Werner}, {Roellig}, {Low}, {Rieke}, {Rieke},
  {Hoffmann}, {Young}, {Houck}, {Brandl}, {Fazio}, {Hora}, {Gehrz}, {Helou},
  {Soifer}, {Stauffer}, {Keene}, {Eisenhardt}, {Gallagher}, {Gautier}, {Irace},
  {Lawrence}, {Simmons}, {Van Cleve}, {Jura}, {Wright}, \&
  {Cruikshank}}]{Werner2004}
{Werner}, M.~W., {Roellig}, T.~L., {Low}, F.~J., {et~al.} 2004, \apjs, 154, 1

\bibitem[{{Willacy} {et~al.}(1998){Willacy}, {Langer}, \&
  {Velusamy}}]{Willacy1998}
{Willacy}, K., {Langer}, W.~D., \& {Velusamy}, T. 1998, \apjl, 507, L171

\bibitem[{{Wolf}(1923)}]{Wolf1923}
{Wolf}, M. 1923, Astronomische Nachrichten, 219, 109

\bibitem[{{Ysard} {et~al.}(2013){Ysard}, {Abergel}, {Ristorcelli}, {Juvela},
  {Pagani}, {K{\"o}nyves}, {Spencer}, {White}, \& {Zavagno}}]{Ysard2013}
{Ysard}, N., {Abergel}, A., {Ristorcelli}, I., {et~al.} 2013, \aap, 559, A133

\bibitem[{{Zucconi} {et~al.}(2001){Zucconi}, {Walmsley}, \&
  {Galli}}]{Zucconi2001}
{Zucconi}, A., {Walmsley}, C.~M., \& {Galli}, D. 2001, \aap, 376, 650

\end{thebibliography}
 \Online
 \begin{appendix}
 \section{Multi-layered picture}
\begin{figure*}[t]
\centering
\includegraphics[width=0.9\linewidth]{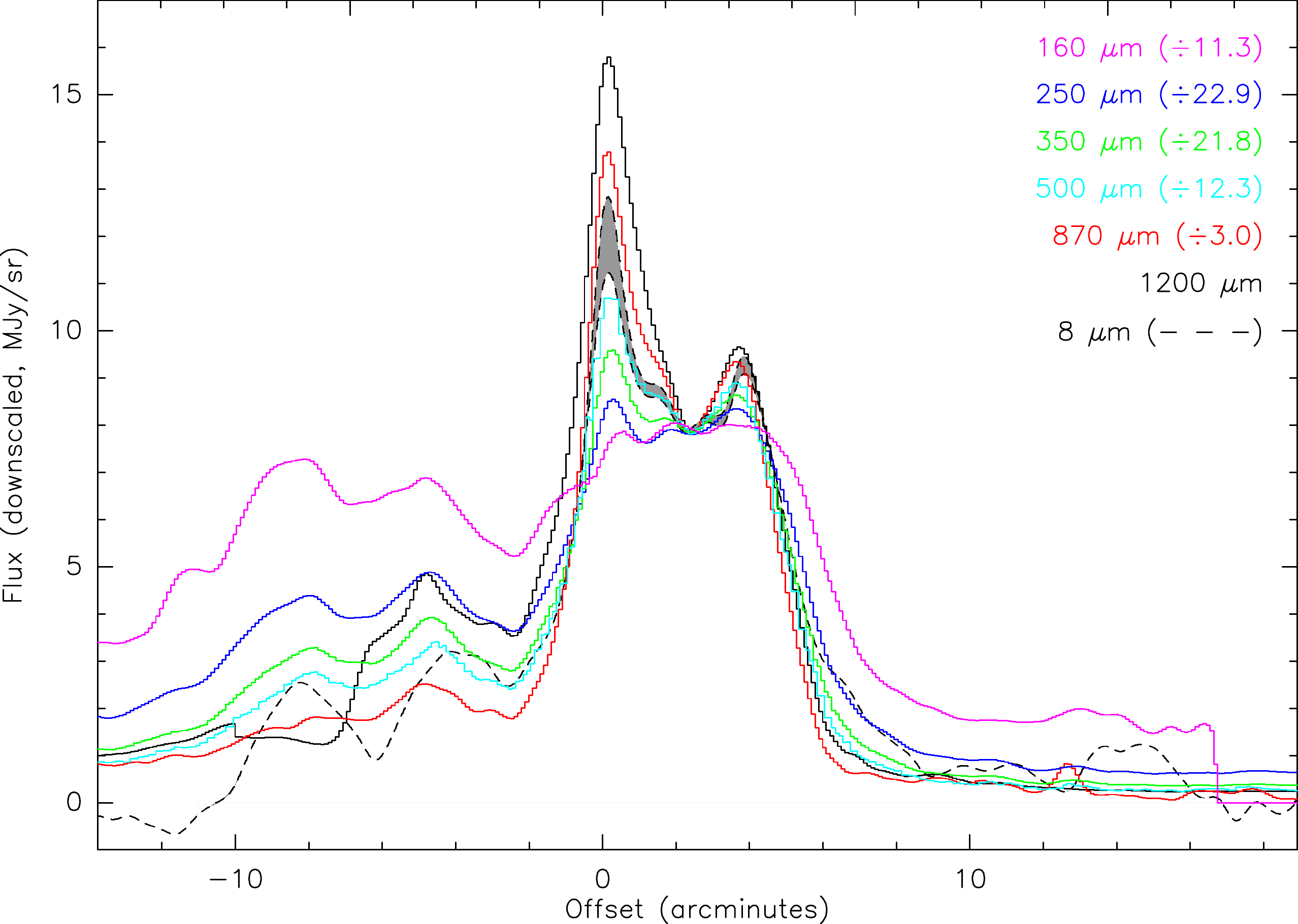}
\caption{L183 cut for dust emission. It includes data from \textit{Herschel} PACS and SPIRE, APEX/LABOCA, and IRAM-30m/MAMBO data, all convolved to 37\arcsec\ resolution. The fluxes are scaled down by the amount indicated to the right of the wavelengths to align all the fluxes on the saddle point. The 8 \mic\ opacity range is displayed (dashed lines filled in  grey), the 100 \mic\ cut is omitted since it only traces the cloud content at its surface. This is the developed version of Fig. \ref{fig:L183_Emission}. The colour  is changed for each wavelength to help separate them.}
\label{fig:App_L183_Emission}
\end{figure*}
 \begin{figure*}[b!]
\centering
 \includegraphics[width=\linewidth]{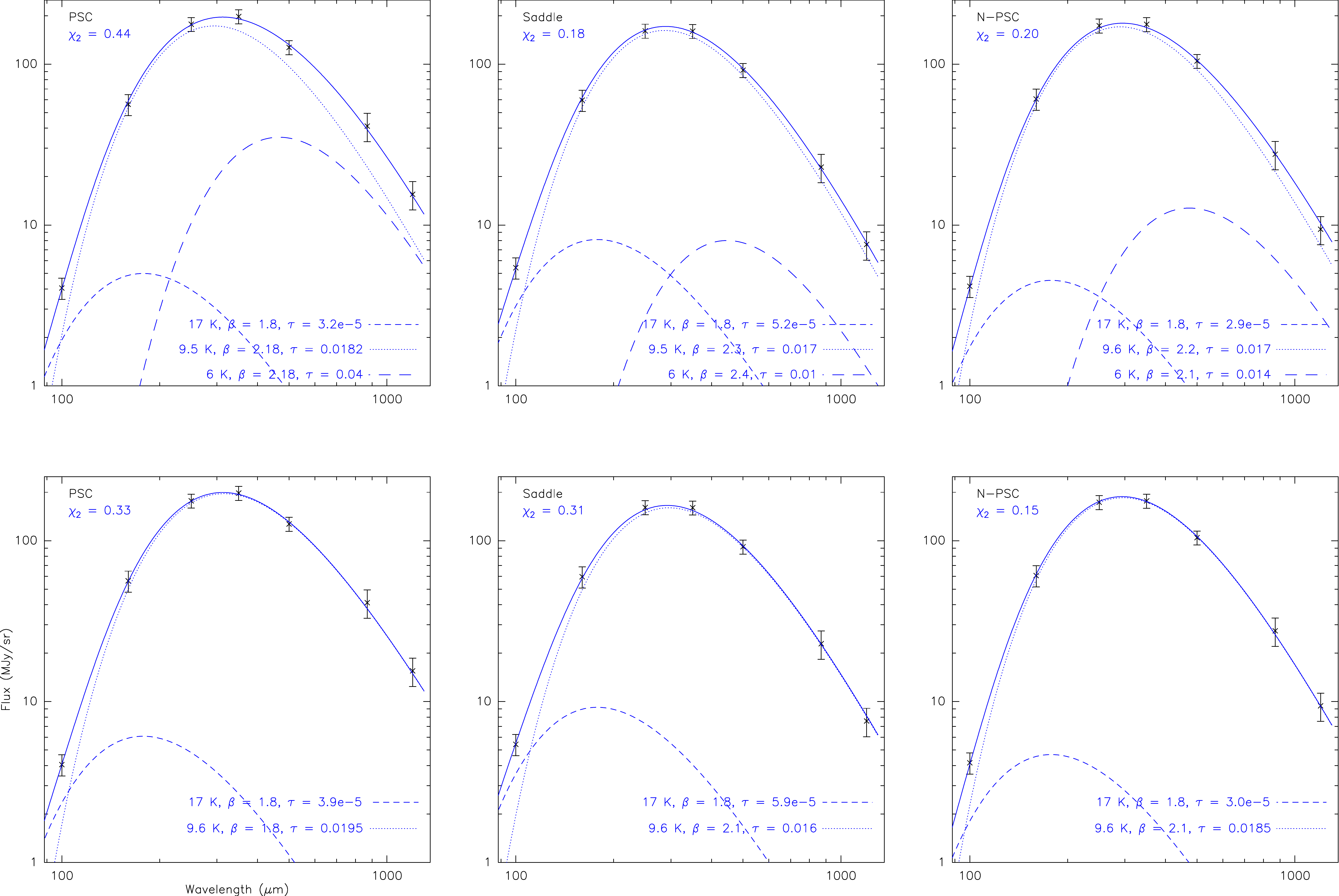}
 \caption{SEDs of the three points of interest defined in Fig. \ref{fig:L183_8_250_870um_3y} along the cut (from left to right, the main PSC, the saddle and the northern PSC). Top row, the SEDs are fitted with three modified blackbodies, bottom row, they are fitted with two modified blackbodies. Opacity is given at 300 \mic. This is the developed version of Fig. \ref{fig:L183_6_SEDs}.} 
 \label{fig:App_L183_6_SEDs}
 \end{figure*}
 Figure \ref{fig:App_L183_Emission} shows the  layers of Fig. \ref{fig:L183_Emission} assembled together for direct comparison, for pdf viewers not understanding Javascript, and for printers.
 Similarly, Fig. \ref{fig:App_L183_6_SEDs} shows the  layers of Fig. \ref{fig:L183_6_SEDs} (the three positions tested with two or three blackbody components) side by side.
 \section{Institutional acknowledgements}\label{AppB}
 This work is based on observations carried out with the IRAM 30m Telescope. IRAM is supported by INSU/CNRS (France), MPG (Germany), and IGN (Spain) and on data acquired with the Atacama Pathfinder Experiment (APEX). APEX is a collaboration between the Max-Planck-Institut f\"ur Radioastronomie, the European Southern Observatory, and the Onsala Space Observatory. 
 %
    {\it Planck} \emph{(http://www.esa.int/Planck)} is a project of the European Space Agency -- ESA -- with instruments provided by two scientific consortia funded
    by ESA member states (in particular the lead countries: France and Italy) with
    contributions from NASA (USA), and telescope reflectors provided in a
    collaboration between ESA and a scientific Consortium led and funded by
    Denmark. 
 %
 {\it Herschel} is an ESA space observatory with science instruments provided
    by European-led Principal Investigator consortia and with important
    participation from NASA. 
 
 \end{appendix}

\end{document}